\documentclass[manuscript,screen]{acmart}

\usepackage{listings}
\usepackage{xcolor}

\newif\ifproposal
    \proposalfalse

\newif\ifthesis
    \thesisfalse

\newif\ifieeejournal
    \ieeejournalfalse
\newif\ifacmjournal
    \acmjournalfalse
\newif\ifbiographies
    \biographiesfalse
\newif\ifreplytoreviewers
    \replytoreviewersfalse

\newif\ifieeeconference
    \ieeeconferencefalse

\newif\ifisf
    \isffalse

\newif\ifmicro
    \microfalse

    \acmjournaltrue
    \biographiestrue
    \replytoreviewersfalse

\usepackage{amsmath}
\usepackage{multirow}
\usepackage{subfigure}
\usepackage{etex}
\usepackage{basic_packages}
\renewcommand{\eqref}[1]{(\ref{#1})}
\newcommand{\secref}[1]{\mbox{Section~\ref{#1}}}

\newcommand{\figref}[1]{\mbox{Fig.~\ref{#1}}}

\newcommand{\tblref}[1]{\mbox{Table~\ref{#1}}}

\definecolor{applegreen}{rgb}{0.55, 0.71, 0.0}

\usepackage{environ}
\newcounter{charcount}
\newcounter{charlim}
\NewEnviron{countenv}[2][]{%
  \setcounter{charcount}{0}%
  \expandafter\countem\BODY\relax\EOE
  \relax%
  \ifnum\value{charcount}<\value{charlim}\relax
      \begin{#2}\BODY\end{#2}
  \else
      \begin{#2}#1 \BODY  
         \color{red}\thecharcount{} letters in the #2 exceeds
         \thecharlim{} character limit. 
         Use the ``charlim'' parameter to change the character limit. \color{black}
       \end{#2} 
  \fi
}

\long\def\countem#1#2\EOE{%
  \stepcounter{charcount}%
  \ifx\relax#2
    \def\next{\relax}%
  \else
    \def\next{\countem#2\EOE}%
  \fi
  \expandafter\next%
}

\usepackage{scalerel}
\usetikzlibrary{svg.path}
\definecolor{orcidlogocol}{HTML}{A6CE39}
\tikzset{
  orcidlogo/.pic={
    \fill[orcidlogocol] svg{M256,128c0,70.7-57.3,128-128,128C57.3,256,0,198.7,0,128C0,57.3,57.3,0,128,0C198.7,0,256,57.3,256,128z};
    \fill[white] svg{M86.3,186.2H70.9V79.1h15.4v48.4V186.2z}
                 svg{M108.9,79.1h41.6c39.6,0,57,28.3,57,53.6c0,27.5-21.5,53.6-56.8,53.6h-41.8V79.1z M124.3,172.4h24.5c34.9,0,42.9-26.5,42.9-39.7c0-21.5-13.7-39.7-43.7-39.7h-23.7V172.4z}
                 svg{M88.7,56.8c0,5.5-4.5,10.1-10.1,10.1c-5.6,0-10.1-4.6-10.1-10.1c0-5.6,4.5-10.1,10.1-10.1C84.2,46.7,88.7,51.3,88.7,56.8z};
  }
}

\newcommand\orcidicon[1]{\href{https://orcid.org/#1}{\mbox{\scalerel*{
\begin{tikzpicture}[yscale=-1,transform shape]
\pic{orcidlogo};
\end{tikzpicture}
}{|}}}}

\newcommand{\X}{$\times$\xspace}    
    
\newcommand{\sota}{SOTA\xspace}

\newcommand{\mm}{\,\si{\mm}\xspace}

\newcommand{\m}{\,\si{\m}\xspace}

\newcommand{\nm}{\,\si{\nm}\xspace}

\newcommand{\um}{\,\si{\um}\xspace}

\newcommand{\s}{\,\si{\s}\xspace}
\newcommand{\ps}{\,\si{\ps}\xspace}
\newcommand{\ns}{\,\si{\ns}\xspace}
\newcommand{\us}{\,\si{\us}\xspace}
\newcommand{\ms}{\,\si{\ms}\xspace}

\newcommand{\bit}{\,\si{\bit}\xspace}

\newcommand{\GB}{\,\si{\giga\byte}\xspace}

\newcommand{\Hz}{\,\si{\Hz}\xspace}
\newcommand{\kHz}{\,\si{\kHz}\xspace}
\newcommand{\MHz}{\,\si{\MHz}\xspace}
\newcommand{\GHz}{\,\si{\GHz}\xspace}

\newcommand{\W}{\,\si{\watt}\xspace}
\newcommand{\mW}{\,\si{\milli\watt}\xspace}
\newcommand{\uW}{\,\si{\micro\watt}\xspace}

\newacronym{iot}{IoT}{internet-of-things}

\newacronym{snr}{SNR}{signal-to-noise ratio}
    
\newacronym{ppa}{PPA}{performance, power, and area}

\newacronym{cdf}{CDF}{cumulative distribution function}
\newacronym{pdf}{PDF}{probabililty distribution function}
\newacronym{ip}{IP}{intellectual property}
\newacronym{fir}{FIR}{finite impulse response}
\newacronym{dsp}{DSP}{digital signal processing}
\newacronym{lut}{LUT}{look-up table}
\newacronym{mac}{MAC}{multiply-and-accumulate}
    \newcommand{\mac}{\gls{mac}\xspace}
    \newcommand{\macs}{\glspl{mac}\xspace}

\newacronym{dl}{DL}{deep learning}
    
\newacronym{ml}{ML}{machine learning}
    
\newacronym{ai}{AI}{artificial-intelligence}

\newacronym{cnn}{CNN}{}
    \newcommand{\cnn}{\gls{cnn}\xspace}
\newacronym{dnn}{DNN}{deep neural network}
    \newcommand{\dnn}{\gls{dnn}\xspace}
    \newcommand{\dnns}{\glspl{dnn}\xspace}
\newacronym[longplural={graphic processing units}]{gpu}{GPU}{graphic processing unit}

\newacronym[longplural={central processing units}]{cpu}{CPU}{central processing unit}

\newacronym{tpu}{TPU}{tensor processing unit}
    
\newacronym{relu}{ReLU}{}
    \newcommand{\relu}{\gls{relu}\xspace}
\newacronym{llm}{LLM}{Large Language Model}

\newacronym{fpga}{FPGA}{}
    \newcommand{\fpga}{\gls{fpga}\xspace}
\newacronym{enics}{EnICS}{Emerging NanoScaled Integrated Circuits \& Systems}

\newacronym{BIU}{BIU}{Bar-Ilan University\xspace}
     
\newacronym{UNICAL}{UNICAL}{University of Calabria\xspace}
    
\newacronym{DIMES}{DIMES}{Department of Computer Engineering, Modeling, Electronics and Systems\xspace}
    
\newacronym{USFQ}{USFQ}{Universidad San Francisco de Quito\xspace}
         
\newacronym{EPFL}{EPFL}{\'Ecole Polytechnique F\'ed\'erale de Lausanne\xspace}
     
\newacronym{isf}{ISF}{Israel Science Fund\xspace}
\newacronym{iia}{IIA}{Israel Innovation Authority\xspace}

\newacronym{itrs}{ITRS}{International Technology Roadmap for Semiconductors}
\newacronym{vlsi}{VLSI}{very large scale integration}
\newacronym{asic}{ASIC}{application specific integrated circuit}
\newacronym{pcb}{PCB}{printed circuit board}
\newacronym{cmos}{CMOS}{complementary-metal-oxide-semiconductor}

\newacronym[longplural={systems-on-chip}]{soc}{SoC}{system-on-chip}

\newacronym[longplural={integrated circuits}]{ic}{IC}{integrated circuit}

\newacronym{mc}{MC}{Monte Carlo}
     
\newacronym{mep}{MEP}{minimum energy point}

\newacronym[longplural={non-volatile memories}]{nvm}{NVM}{non-volatile memory}

\newacronym{vdd}{$V_{\text{DD}}$}{supply voltage}
    
\newacronym{gnd}{$GND$}{ground}
    
\newacronym{subvt}{sub-$V_{\text{T}}$}{sub-threshold}
    
\newacronym{nearvt}{near-$V_{\text{T}}$}{near threshold}
\newacronym{vt}{$V_{\text{T}}$}{threshold voltage}

\newacronym{vgs}{$V_{\text{GS}}$}{gate-to-source voltage} 
\newacronym{vds}{$V_{\text{DS}}$}{drain-to-source voltage} 
\newacronym{vbs}{$V_{\text{BS}}$}{body-to-source voltage} 
\newacronym{vgb}{$V_{\text{GB}}$}{gate-to-body voltage} 
\newacronym{dibl}{DIBL}{drain induced barrier lowering}
\newacronym{gidl}{GIDL}{gate induced drain leakage} 
\newacronym{ids}{$I_{\text{DS}}$}{drain-to-source current}
\newacronym{sce}{SCE}{short channel effect}
\newacronym{rsce}{RSCE}{reverse short channel effect}
\newacronym{tox}{$t_{\text{ox}}$}{gate oxide thickness}
\newacronym{L}{$L$}{channel length}
\newacronym{W}{$W$}{channel width}
\newacronym{rbb}{RBB}{reverse body biasing}
\newacronym{fbb}{FBB}{forward body biasing}
\newacronym{btbt}{BTBT}{band-to-band tunneling}
\newacronym{bjt}{BJT}{bipolar junction transistor}
\newacronym{hvt}{HVT}{high threshold voltage}
\newacronym{lvt}{LVT}{low threshold voltage}
\newacronym{nvt}{NVT}{nominal threshold voltage}
\newacronym{pmos}{PMOS}{p-type MOSFET}
\newacronym{nmos}{NMOS}{n-type MOSFET}
\newacronym{isub}{$I_\text{sub}$}{sub-threshold leakage}
\newacronym{igate}{$I_\text{gate}$}{gate leakage}
\newacronym{ibulk}{$I_\text{bulk}$}{bulk leakage}
\newacronym{vbb}{$V_{\text{BB}}$}{body voltage} 
    
\newacronym{ptm}{PTM}{predictive technology model}

\newacronym{pdk}{PDK}{process design kit}

\newacronym{sc}{SC}{standard cell}
\newacronym{vtc}{VTC}{voltage transfer characteristic}
\newacronym{dff}{DFF}{Data Flip-Flop}
\newacronym{dcvsl}{DCVSL}{differential cascade voltage switch logic}

\newacronym{dif}{DIF}{digital implementation flow}
\newacronym{hdl}{HDL}{hardware description language}
\newacronym{rtl}{RTL}{register transfer level}
\newacronym{eda}{EDA}{electronic design automation}
\newacronym{cad}{CAD}{computer-aided design}
\newacronym{pr}{P\&R}{place and route}
\newacronym{cts}{CTS}{clock-tree synthesis}
\newacronym{sta}{STA}{static timing analysis}
\newacronym{edi}{EDI}{Cadence Encounter Design Implementation}
\newacronym{s_dc}{DC}{Synopsys Design Compiler}
\newacronym{sdc}{SDC}{Synopsys Design Constraints}
\newacronym{vcd}{VCD}{value change dump}
\newacronym{pvt}{PVT}{Process-Voltage-Temperature}
\newacronym{scm}{SCM}{standard cell memory}
    
\newacronym{ips}{IPS}{instructions per second}

\newacronym{eflash}{eFlash}{embedded Flash}

\newacronym[longplural={Storage Class Memories}]{scmems}{SCM}{Storage Class Memory}

\newacronym{ddr}{DDR}{dual-data rate}

\newacronym{sata}{SATA}{Serial Advanced Technology Attachment}
    
\newacronym{nvme}{NVMe}{Non-Volatile Memory Express}
    
\newacronym{pcie}{PCIe}{Peripheral Component Interconnect Express}
     
\newacronym[longplural={hard-Disk drives}]{hdd}{HDD}{hard-Disk drive}

\newacronym[longplural={solid-State drives}]{ssd}{SSD}{solid-State drive}

\newacronym[longplural={high-bandwidth memories}]{hbm}{HBM}{high-bandwidth memory}

\newacronym[longplural={dual-inline memory modules}]{dimm}{DIMM}{dual-inline memory module}

\newacronym[longplural={dynamic random-access memories}]{dram}{DRAM}{dynamic random-access memory}

\newacronym{fifo}{FIFO}{first-in first-out}
\newacronym{lifo}{LIFO}{last-in first-out}
\newacronym[longplural={content addressable memories}]{cam}{CAM}{content addressable memory}
    
\newacronym{L1}{L1}{level-1}
\newacronym{L2}{L2}{level-2}
\newacronym{L3}{L3}{level-3}
\newacronym{L4}{L4}{level-4}
\newacronym{simd}{SIMD}{single-instruction multiple-data}

\newacronym{bc}{BC}{bitcell}

\newacronym{bl}{BL}{bitline}

\newacronym{sln}{SL}{sourceline}

\newacronym{wl}{WL}{wordline}

\newacronym[longplural={Gain-Cell embedded DRAMs}]{gcedram}{GC-eDRAM}{Gain-Cell embedded DRAM}

\newacronym{sixt}{6T}{6-transistor}
    
\newacronym[longplural={static random-access memories}]{sram}{SRAM}{static random-access memory}

\newacronym[longplural={six-transistor static random access memories}]{sixtsram}{6T-SRAM}{six-transistor static random access memory}

\newacronym[longplural={embedded DRAMs}]{edram}{eDRAM}{embedded DRAM}

\newacronym[longplural={multi-level cells}]{mlc}{MLC}{multi-level cell}

\newacronym{mw}{MW}{write transistor}
\newacronym{mr}{MR}{read transistor}
\newacronym{sn}{SN}{storage node}

\newacronym{wwl}{WWL}{write word line}

\newacronym{rwl}{RWL}{read word line}

\newacronym{sa}{SA}{sense amplifier}
\newacronym{drv}{DRV}{data retention voltage}
\newacronym{nwl}{NWL}{negative word line}
\newacronym{bist}{BIST}{built-in self-test}
\newacronym{bisr}{BISR}{built-in self-repair}
\newacronym{ecc}{ECC}{error correction code}
\newacronym{snm}{SNM}{static noise margin}
\newacronym{rsnm}{RSNM}{read static noise margin}
\newacronym{wsnm}{WSNM}{write static noise margin}
\newacronym{dnm}{DNM}{dynamic noise margin}
\newacronym{drt}{DRT}{data retention time}

\newacronym{lrs}{LRS}{low resistance state}

\newacronym{hrs}{HRS}{high resistance state}

\newacronym[longplural={phase-change memories}]{pcm}{PCM}{phase-change memory}

\newacronym[longplural={resistive RAMs}]{rram}{RRAM}{resistive RAM}

\newacronym{stt}{STT}{spin-transfer torque}

\newacronym[longplural={spin-transfer torque magnetic random-access memories}]{sttmram}{STT-MRAM}{spin-transfer torque magnetic random-access memory}

\newacronym[longplural={magnetic random-access memories}]{mram}{MRAM}{magnetic random-access memory}

\newacronym{mtj}{MTJ}{magnetic tunnel junction}

\newacronym{smtj}{SMTJ}{single-barrier MTJ}

\newacronym{dmtj}{DMTJ}{double-barrier MTJ}

\newacronym{mim}{MIM}{metal-insulator-metal}

\newacronym{euv}{EUV}{extreme ultra-violet}
     
\newacronym{soi}{SOI}{silicon-on-insulator}
\newacronym{fdsoi}{FD-SOI}{fully-depleted silicon-on-insulator}
    
\newacronym{rdf}{RDF}{random dopant fluctuations}
\newacronym{ocv}{OCV}{on-chip variation}

\newacronym{lpa}{LPA}{Leakage Power Analysis}
\newacronym{dpa}{DPA}{Differential Power Analysis}
\newacronym{puf}{PUF}{Physical Unclonable Function}

\newacronym{ser}{SER}{soft errors}
     
\newacronym{seu}{SEU}{single-event upset}
     
\newacronym{qcrit}{$Q_\text{crit}$}{critical charge}
\newacronym{tmr}{TMR}{triple modular redundancy}
\newacronym{dmr}{DMR}{dual modular redundancy}
\newacronym{edac}{EDAC}{error detection and correction}
\newacronym{secded}{SECDED}{single error correction~-- double error detection}
\newacronym{dected}{DECTED}{double error correction~-- triple error detection}

\newacronym{smu}{SMU}{source/measure unit}
\newacronym{dmm}{DMM}{digital multimeter}
\newacronym{AF}{AF}{}
    \newcommand{\AF}{\gls{AF}\xspace}
    \newcommand{\AFs}{\glspl{AF}\xspace}

\newacronym{cordic}{CORDIC}{COordinate Rotation DIgital Computer}
    \newcommand{\cordic}{\gls{cordic}\xspace}

\newcommand{\softmax}{SoftMax\xspace}
\newcommand{\selu}{SeLU\xspace}
\newcommand{\gelu}{GeLU\xspace}

\newacronym{fsm}{FSM}{finite state machine}
    \newcommand{\fsm}{\gls{fsm}\xspace}

\newacronym{PE}{PE}{}
    \newcommand{\PE}{\gls{PE}\xspace}
    \newcommand{\PEs}{\glspl{PE}\xspace}

\newacronym{RNN}{RNN}{recurrent neural networks}
    \newcommand{\RNN}{\gls{RNN}\xspace}
    \newcommand{\RNNs}{\glspl{RNN}\xspace}

\usepackage{clipboard} 
\usepackage{subfiles} 
\usepackage{soul} 

\newcounter{reviewer}
\newcounter{point}[reviewer]
\renewcommand{\thepoint}{P\,\thereviewer.\arabic{point}} 

\newenvironment{reply}
   {\color{blue} \medskip \textbf{Reply}:\  }
   {\color{black} \medskip }

\newcommand{\shortreply}[2][]{\color{blue} \medskip \noindent \begin{sf}\textbf{Reply}:\  #2
	\ifthenelse{\equal{#1}{}}{}{ \hfill \footnotesize (#1)}%
	\color{black} \medskip \end{sf}}

 \newenvironment{changes}
   {\color{blue} \medskip \textbf{Authors' Action}:\  }
   {\color{black} \medskip }

\newclipboard{output-ACM_Journal} 

\setcopyright{none}
\copyrightyear{2024}
\acmYear{2024}
\acmDOI{XXXXXXX.XXXXXXX}

\begin{document}

\title{\Copy{CORDIC Is All You Need}{CORDIC Is All You Need}}

\author{Omkar Kokane}
\affiliation{%
  \institution{NSDCS Lab, Indian Institute of Technology Indore}
  \city{Indore}
  \country{India}
}
\email{mt2302102023@iiti.ac.in}
\orcid{0009-0000-6288-7231}

\author{Adam Teman}

\affiliation{%
  \institution{EnICS Labs, Bar Ilan University}
  \city{Ramat Gan}
  \country{Israel}
}
\email{adam.teman@biu.ac.il}
\orcid{0000-0002-8233-4711}

\author{Anushka Jha}
\authornote{These authors contributed equally to this research.}
\email{ee220002013@iiti.ac.in}
\orcid{1234-5678-9012}
\author{Guru Prasath SL}
\authornotemark[1]
\email{ee230002026@iiti.ac.in}
\orcid{1234-5678-9012}
\affiliation{%
  \institution{Indian Institute of Technology Indore}
  \city{Indore}
  \country{India}
}

\author{Gopal Raut}
\authornote{These authors contributed equally to this research.}
\email{gopalraut05@gmail.com}
\orcid{0000-0002-1046-9457}
\author{Mukul Lokhande}
\authornotemark[2]
\email{phd2201102020@iiti.ac.in}
\orcid{0009-0001-8903-5159}
\affiliation{%
  \institution{NSDCS Lab, Indian Institute of Technology Indore}
  \city{Indore}
  \country{India}
}

\author{S. V. Jaya Chand}
\authornotemark[1]
\email{ee220002076@iiti.ac.in}
\orcid{0009-0007-6855-7859}
\author{Tanushree Dewangan}
\authornotemark[1]
\email{ee220002077@iiti.ac.in}
\orcid{0009-0009-3889-1228}
\affiliation{%
  \institution{Indian Institute of Technology Indore}
  \city{Indore}
  \country{India}
}

\author{Santosh Kumar Vishvakarma}
\affiliation{%
  \institution{NSDCS Lab, Indian Institute of Technology Indore}
  \city{Indore}
  \country{India}
}
\email{skvishvakarma@iiti.ac.in}
\orcid{0000-0003-4223-0077}



\renewcommand{\shortauthors}{Kokane et al.}

\begin{abstract}

Artificial intelligence necessitates adaptable hardware accelerators for efficient high-throughput million operations. We present pipelined architecture with CORDIC block for linear MAC computations and nonlinear iterative Activation Functions (\textbf{AF}) such as $\tanh$, $sigmoid$, and \softmax. This approach focuses on a Reconfigurable Processing Engine (\textbf{RPE}) based systolic array, with 40\% pruning rate, enhanced throughput up to 4.64$\times$, and reduction in power and area by 5.02 $\times$ and 4.06 $\times$ at CMOS 28 nm, with minor accuracy loss. FPGA implementation achieves a reduction of up to 2.5 $\times$ resource savings and 3 $\times$ power compared to prior works. The Systolic CORDIC engine for Reconfigurability and Enhanced throughput (SYCore) deploys an output stationary dataflow with the CAESAR control engine for diverse AI workloads such as Transformers, RNNs/LSTMs, and DNNs for applications like image detection, LLMs, and speech recognition. The energy-efficient and flexible approach extends the enhanced approach for edge AI accelerators supporting emerging workloads.

\end{abstract}



\received{XX Mon 20XX}
\received[revised]{XX Mon 20XX}
\received[accepted]{XX Mon 20XX}

\maketitle

\section{Introduction}
\label{sec_introduction}

Artificial Intelligence and Machine Learning (\textbf{AI/ML}) have been extensively implemented and embraced across various technological domains. Among the myriad algorithmic frameworks, \textbf{\dnns} have captured significant attention due to their vast applicability in fields like science, engineering, agriculture, healthcare, etc. DNNs have revolutionized image classification, audio recognition, and natural language processing ~\cite{raut2023designingperformacecentricMAC}.
 \dnns are structured with multiple layers, starting from the input, passing through several hidden layers, and concluding with the output layers. During inference, these layers form a directed acyclic graph with nodes and edges leading to neurons in subsequent layers. Each node, or neuron, executes two core tasks: First, it employs a Multiply, Add and Accumulate( \textbf{\mac}) unit to aggregate the weighted input features and second, it applies the Activation Function(\textbf{\AF}) to the resulting sum. 
 A deep neural network (DNN) is a type of Artificial Neural Network(\textbf{ANN}) that consists of multiple hidden layers between the input and output, typically structured as either a Convolutional Neural Network (\textbf{\cnn}) or Fully Connected (\textbf{FC}). Researchers have focused on developing specialized hardware solutions, including Application-Specific Integrated Circuits (\textbf{ASICs}), Graphics Processing Units (\textbf{GPUs}), Field-Programmable Gate Arrays (\textbf{\fpga}), and multi-threaded CPUs~\cite{Zhu-16bSoftMax-TCASII20-FXp_SoftMax, FxP8-bSoftMax_APCCAS18, MRao-ISQED24-Floating-point-SoftMax}, to enhance the efficiency of \dnn implementations.

 ASICs are specialized hardware optimized for a specific task, sacrificing the adaptability of a reconfigurable architecture for improved power efficiency compared to platforms such as FPGAs and GPUs. These latter platforms, while more versatile and capable of executing a variety of tasks, tend to have higher power consumption~\cite{Resource-constrain-Architectural-paper-over-FPGA, ASIC-and-FPGA-analysis-for-CNN}. FPGAs provides cost efficiency, and flexibility, with limited on-chip memory and resources~\cite{ASIC-and-FPGA-analysis-for-CNN}. Thus, innovative design methodologies are required to manage the area, complexity, and reconfigurability challenges in \dnns. Additionally, a \dnn comprises various layers, including CNN, feed-forward, normalization, long-term short-term memory (\textbf{LSTM)}, Gated Recurrent Unit (\textbf{GRU}), pooling, and activation layers like Rectified Linear Unit(\textbf{\relu}), $\tanh$, and sigmoid~\cite{Zhu-16bSoftMax-TCASII20-FXp_SoftMax, Kokane_2024-SAMRT-AF}. The core components of a \dnn model are the Processing Engines (PEs), or artificial neurons. These involve multiplication and accumulation processes followed by a nonlinear \AF. Multiplication and accumulation operations, handled by \mac units through complex adder and compressor design, are computationally intensive. The output is then subjected to nonlinear transformations using \AFs such as sigmoid, $\tanh$, and \relu, commonly employed in \dnns~\cite{raut2023empirical-ACM}. Activation functions are crucial in \dnn architectures to enable features like model size reduction and prevention of overfitting issues~\cite{han2015deep}. Various methods, including LookUp-Tables (\textbf{LUT}), \cordic, and PieceWise Linear (\textbf{PWL}), can implement these Activation Functions. While some designs are resource-intensive in terms of area and power, the \cordic design is praised for its area and power efficiency; however, a considerable limitation is its low throughput.

\begin{figure}[h]
    \centering
    \begin{minipage}{0.40\textwidth}
        \centering
        \includegraphics[width=\linewidth,height =50mm]{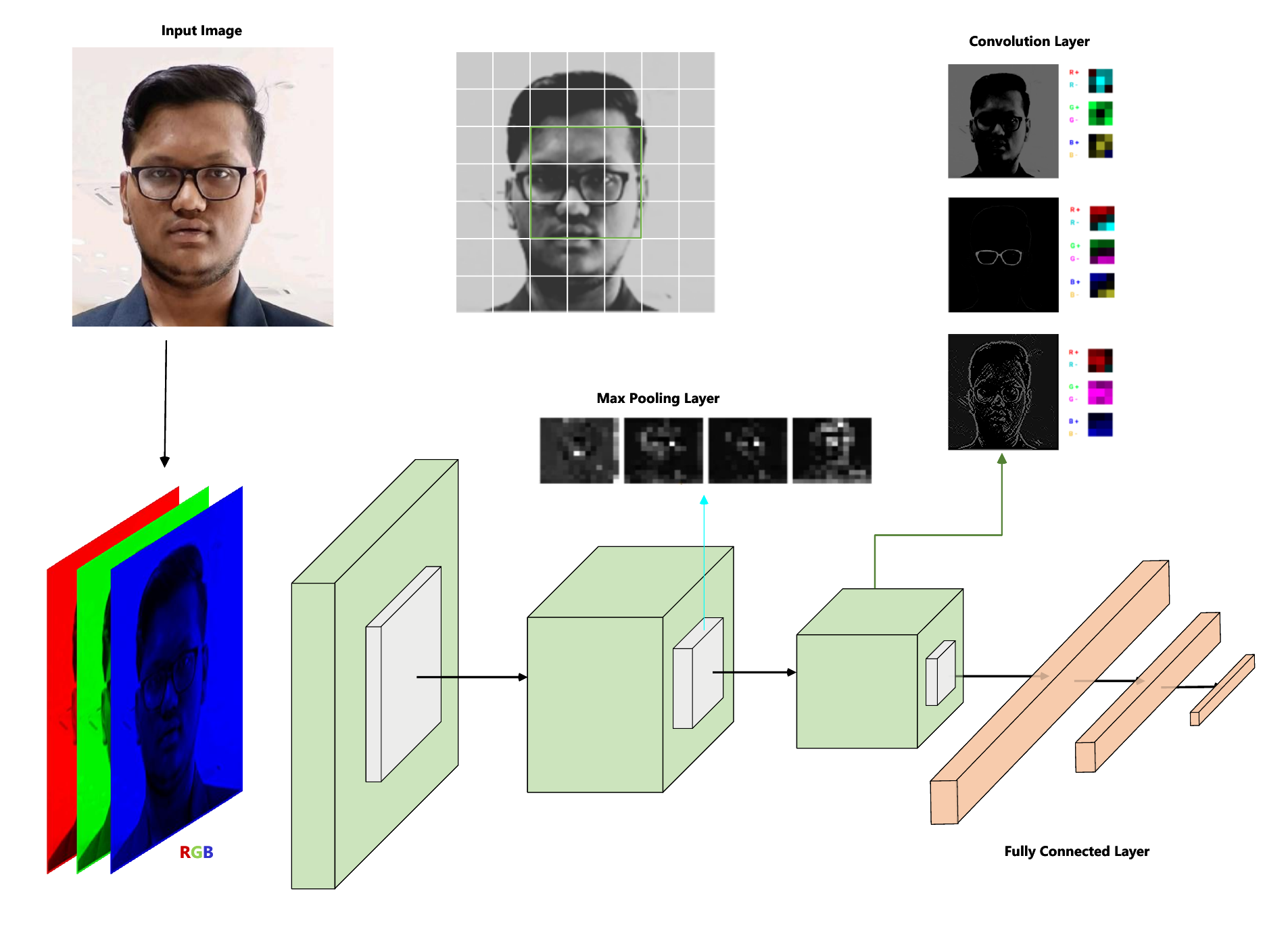}
        \label{fifig:intro_DNN}
    \end{minipage}
    \hfill
    \begin{minipage}{0.59\textwidth}
        \centering
        \includegraphics[width=\linewidth,height =50mm]{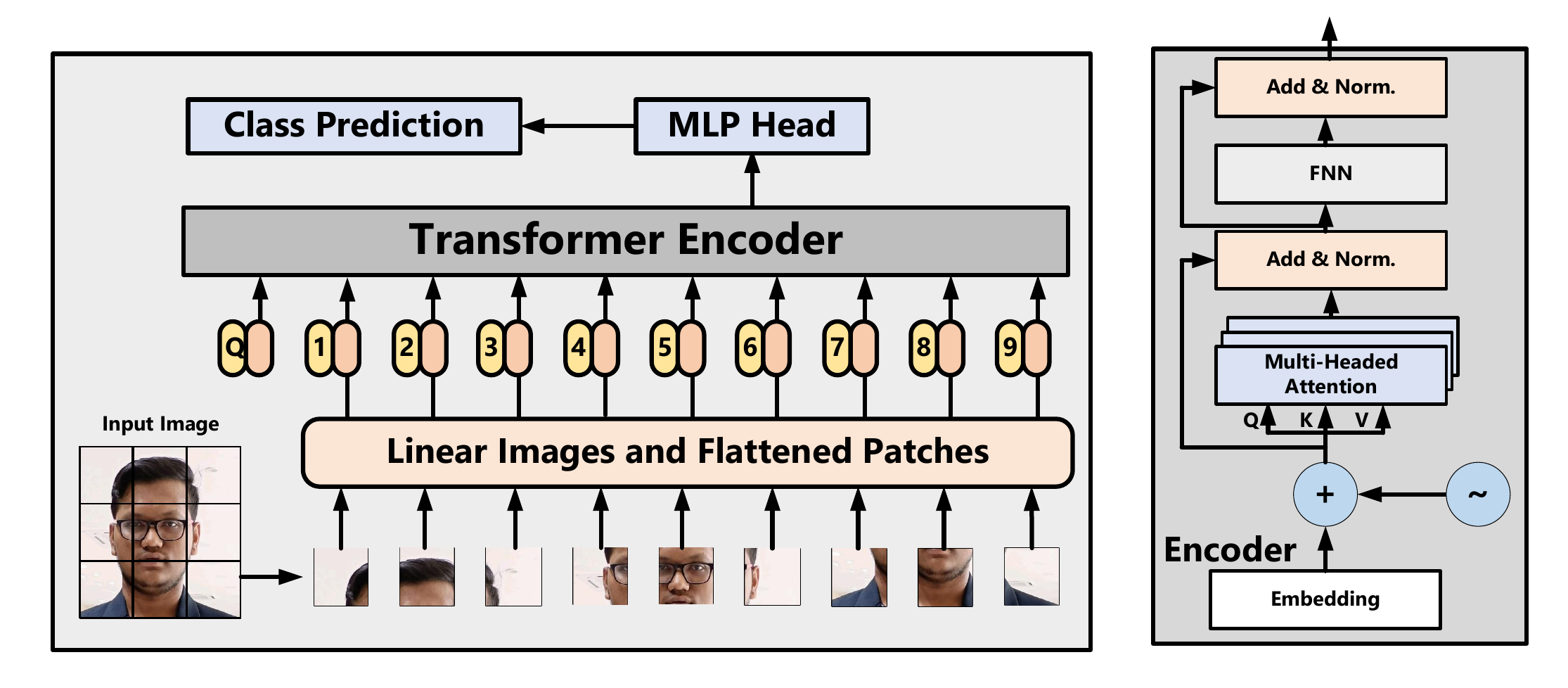}
        \label{fig:intro_transformers}
    \end{minipage}
    \caption{Typical demonstration of (a) \dnn and (b) Transformers model, depicted intermediate layers such as Conv, Pooling, MHA, and FFNN}

    \vspace{-5mm}
\end{figure}
Prior AI accelerator designs prioritize area and power efficiency and focus on the \cordic computational algorithm ~\cite{Unified-CORDIC, raut2020cordic, Flex-PE, Kokane_2024-SAMRT-AF}. This design comprises fundamental logic components, including a shift register, adder/subtractor circuits, multiplexers, and some memory elements that facilitate a range of arithmetic operations while consuming less power. \cordic algorithm, introduced by Jack E. Volder in 1959, has the adaptability to conduct several arithmetic operations. Mathematically, it achieves linear convergence with minimal resource use~\cite{raut2023empirical-ACM}. Through shift and add-sub operations, the \cordic architecture can execute numerous tasks, such as trigonometric, hyperbolic, and logarithmic functions via vector rotation~\cite{aggarwal2014reconfigurableReconfigurable-CORDIC-multi-modes-multi-trajectory}. We employ \cordic in rotation mode to achieve the $\tanh$ and sigmoid functions, akin to computing hyperbolic functions as exponential terms. Moreover, \cordic can determine square roots and more. The widely used \softmax function can also be implemented through \cordic hardware, which is crucial for various classification tasks and is predominantly used in transformer architectures and capsule networks.

\subsection{Background on Accelerator Design Techniques}
\label{subsec_Related Works on Accelerator Design Techniques}

Enhancing the efficiency of hardware accelerators involves several techniques to increase throughput while minimizing power usage and latency. These techniques include Quantization, Pruning, Sparsity, Power and Clock Gating. Quantization entails lowering the precision of data from a higher to a lower bit-width, which helps reduce the size of adders and multipliers in both weights and inputs, ultimately decreasing delay and power consumption. This process employs methods such as Truncation and Round-to-earest-even, although such methods can compromise precision, resulting in diminished accuracy on a larger scale. Research on eliminating quantizing errors is available in sources like \cite{CANN-Curable-Approximations-DNNAccelerators, Approximation-error-a-study}. Pruning entails minimizing computational data according to its significance, a method utilized by NVIDIA in their GPUs to enhance throughput \cite{nvidia-H100, Samsung-NPU, Intel-MxCore}. Adjusting lower-valued inputs or weights to zero based on the pruning ratio effectively reduces computational requirements. Pruning coordinates with Sparsity, which converts data matrices into sparse formats by removing zeros. Although pruning can affect accuracy, it can be recovered through model fine-tuning, possibly doubling throughput gains. Some of the Latest design also Performs a Hierarchical pruning-sparsity increasing the ratio up to 3:4 reducing the 75\% computation~\cite{HSS-herarical-structured-sparsity}. Sparsity, while complicating data flow, minimizes redundant calculations. At a more fundamental level, Power and Clock Gating are used to conserve energy by disabling power to unused logic blocks; in power gating, a transistor is placed on the power nets to effectively act as an enable pin. Clock gating reduces switching power loss by controlling the clock's frequency to certain devices. These strategies are integral to the Physical Design process during ASIC Chip Fabrication. Though in these optimization techniques, some are applicable at  PE whereas some are applicable and more optimal in architecture design.

The proliferation of edge devices—from smart sensors, including intelligent cameras and soil quality sensors, to autonomous vehicles and data mining—necessitates efficient AI accelerators capable of handling complex computations. These Edge-AI Accelerators are vital for enabling real-time decision-making while minimizing latency, curbing bandwidth demands, and upholding data privacy. Contrary to cloud-based AI, edge computing reduces dependency on external servers, thereby becoming crucial for applications with stringent latency or privacy requirements. Advances in AI and computation accelerators have been significant, particularly in systolic arrays, \cordic processors, and neural network hardware accelerators. Data flow optimization techniques, such as data reuse and row-wise weight or output stationary flows, enhance throughput and lower computational demands. Enhanced Control Engine (\textbf{CE}) designs, along with design-level improvements in \PE, have contributed significantly. \sota hardware accelerators and high-performance computing (\textbf{HPC}) emphasize performance, albeit with substantial area and power costs, thus favouring cloud environments. 

Conversely, edge devices operate under tight constraints, necessitating architectures that efficiently balance these metrics while maintaining adequate accuracy and minimizing area and power usage. In deep neural networks, each neuron executes a Multiply-and-Accumulate (MAC) operation followed by a nonlinear transformation. Different \AFs serve as nonlinear transformations within the network. In a \dnn, the output precision of a \mac unit determines the input precision for the AF. Typically, this precision is expressed as $2N+K$, where $N$ is the bit-width of the \mac input, and $K$ represents additional overhead bits for accumulation, as illustrated in \ref{fig:conventional_PE}. Enhancements in performance have been achieved by truncating \mac outputs, which allow the use of lower precision \AFs, despite a compromise in accuracy. While these configurations deliver high throughput, they also incur significant area overheads in high-precision architecture implementations. Moreover, these architectures are limited by fixed design constraints, making them unsuitable for accommodating configurable \AF implementations and unsupportive of direct scaling for variable bit-precision scenarios. Implementing high-precision processing elements necessitates considerable hardware resources, leading to an exponential growth in resource utilization. Various techniques have been developed, such as storing \AF values and parameters in look-up tables (LUTs)~\cite{Unified-CORDIC}, using piecewise linear and nonlinear transformations~\cite{vachhani2009efficient}, employing the \cordic algorithm, and approximating the \AF. Each design approach offers distinct advantages. The piecewise-linear (PWL) method integrates several linear vectors to implement non-linear functions efficiently. Utilizing a LUT for the PWL representation of an \AF is optimal for minimizing hardware utilization when the number of neurons is less than the available BRAM blocks in an FPGA circuit.

Systolic array architecture is extensively applied for matrix multiplication and CNN operations, foundational in neural networks and transformers. Legacy architectures like Google's TPU utilize systolic arrays for high-throughput and energy-efficient computations. Designs such as Eyeriss~\cite{chen2017eyeriss} incorporate systolic array concepts to optimize energy efficiency through local memory reuse. This exploitation of data's temporal and spatial locality decreases unnecessary memory accesses. Additionally, it facilitates the transfer of partial sums from one \PE to the next, further economizing memory use and computational cycles. This method is known as data-flow optimization, with weight-stationary or output-stationary data flows enhancing throughput and energy efficiency. Hardware architectural design strategies like layer reuse improve resource allocation by recycling hardware across different neural network layers, as discussed in~\cite{raut2022multiplexed-neurocomputing} with Layer-reuse strategies for area-optimized \dnn hardware. Reusing \PE resources across multiple layers allows sequential execution, minimizing hardware underutilization. Advanced architectures also employ time multiplexing. In time multiplexing unused \PEs in one computation can be repurposed for others, maximizing utilization. The data multiplexing approach also reduces memory buffer needs by loading data directly into \PEs using a demultiplexer.

\cordic blocks excel at executing trigonometric, hyperbolic, and logarithmic functions, which highlights their importance in accelerators. Past research, including~\cite{RECON-CORDICNeuron, Unified-CORDIC, tian2024low-CORDIC-division-root, meher2012cordic-CORDIC-for-rotation, aggarwal2012scale-Hyperbolic-CORDIC-waveform-gen, raut2020cordic, vachhani2009efficient-CORDIC-Arch-low-area-high-throughput}, has demonstrated \cordic's efficiency in reducing power and area during activation computations and other tasks. Nonetheless, many conventional designs cater to a single function or are not reconfigurable, presenting difficulties for varied workloads. Integrating \cordic can effectively enhance \AFs, allowing for calculating commonly used \AFs within the same architecture across diverse demands. Although iterative \cordic procedures can lead to latency issues, limiting high-throughput designs that can be mitigated via pipelining approaches. Neural network accelerators are targeting enhanced throughput, utilizing techniques like sparsity and quantization to enhance Tensor Core performance. Despite these advancements, adapting to resource-constrained conditions and supporting emerging workloads like transformers~\cite{zhoutrans}, \dnns, or \RNNs remains challenging. Sparsity in \dnns greatly reduces computational workload, and efficient hardware support for sparsity involves using compressed data formats to lower computation and optimize memory access patterns to reduce power usage. The complexity in using sparsity arises from its control mechanism and address mapping~\cite{han2015deep, EIE}. 

Current accelerators primarily emphasize enhancing specific computational paradigms, such as CNN or ANN, but generally lack a versatile accelerator capable of efficiently executing CNN, ANN, \RNN, and Transformer models. This pursuit of a runtime-configurable accelerator, especially beneficial for devices like smartphones, AR/VR headsets, or diverse demanding applications, introduces considerable complexity. This inflexibility often results in suboptimal utilization of hardware resources. Our study aims to augment design flexibility. Upon reviewing various architectures, we have identified several techniques adeptly applied in novel designs. The study into Deep Neural Network (DNN) accelerators has yielded numerous hardware architectures designed to enhance performance, power efficiency, and area utilization. Notable accelerators, such as FEATHER~\cite{tong2024feather-tk}, Efficient Inference Engine (EIE)~\cite{EIE}, Sparse Convolutional Neural Network (SCNN)~\cite{slimmerCNN}, and Eyeriss~\cite{chen2017eyeriss}, have achieved substantial improvements by exploiting sparsity, weight pruning, and efficient memory access patterns, among other sophisticated techniques. FEATHER reduces computational demands by implementing low-bit quantization and sparsity-aware computing, thus achieving high throughput with minimal area and power, alongside slight accuracy degradation due to quantization~\cite{tong2024feather-tk}. EIE capitalizes on weight pruning during dense matrix-vector operations, thereby significantly lowering on-chip memory usage and energy consumption while boosting computational throughput~\cite{EIE}. SCNN enhances convolution efficiency by adopting a sparse data-flow paradigm that bypasses zero-valued computations, resulting in notable decreases in unnecessary data fetches and data calls, thereby improving memory transfer power consumption and reducing computational latency~\cite{slimmerCNN}. RECON (Resource-Efficient \cordic-Based Neuron Architecture) focuses on achieving hardware efficiency and functional flexibility using the \cordic algorithm for proficient computation of both linear and non-linear \AFs, making it suitable for diverse neural workloads, whether sparse or dense~\cite{RECON-CORDICNeuron}.  QuantMAC introduces a quantization-enabled Multiply Accumulate (MAC) unit to enhance energy efficiency by dynamically adjusting precision levels according to workload demands, employing Quire arithmetic as an extension~\cite{ashar2024quantmac}. Raut et al. introduce \relu-Centric Design Papers, such as~\cite{raut2022multiplexed-neurocomputing}, which highlight the importance of hardware reuse to support multiple \AFs (\AFs), enhancing efficiency while minimizing area overhead~\cite{raut2023designingperformacecentricMAC}. Additionally, a variety of hardware optimization strategies, such as Booth's algorithm, Wallace tree adder, Vedic multiplier, and error-resilient logarithmic multiplier, are employed to enhance performance in high-performance computing. In near-sensor computing, the prevailing trend is towards approximate hardware circuits, with a slight accuracy trade-off deemed acceptable~\cite{RECON-CORDICNeuron, Kokane_2024-LPRE, HOAA, benchmarking}.

 \AFs design are optimized for accelerators within the library's and articles~\cite{charm2024trets,li2024fiberflex}. Gate-level designs have been developed for multiple \AFs, including Hyperbolic, Tangent, \softmax, Gaussian, Sigmoid, \relu, \gelu, and Binary-Step. Similarly, studies concerning \AFs directed us towards employing \cordic. Previously, the LUT method was utilized to store \AF matrices, while the PWL approach applied linear equations like $y=mx+c$, storing multiple values of \(m\) and \(c\) to map functions to specific non-linear \AFs. Making \AFs reliant on memory results in devices that consume high power~\cite{Unified-CORDIC, aggarwal2015concept-Design-reconfigurable-CORDIC}. To circumvent such power-consuming hardware computations, we adopted the \cordic methodology. This paper presents RPE in the form of \textbf{CORDIC(n,m)}, signifying \(n\) \mac stages and \(m/2\) each for hyperbolic and linear stages of the AF. Our evaluation confirms that five pipeline stages can deliver high throughput without sacrificing accuracy with the help of Pareto analysis. Additionally, the architecture is crafted for fixed-point calculations, featuring a design that minimizes area and power usage, and it is adaptable at both architectural and clock levels.

\subsection{Motivation} 

The primary aim of this study is to devise an area-efficient architecture tailored to AI requirements by leveraging hardware resource reuse to handle a range of operations such as CNN, ANN, \RNN/LSTM, and Transformers. However, significant throughput degradation occurs when \dnn hardware processes Transformers, necessitating advanced mapping techniques~\cite{sinha2007fully-pipelined-CORDIC-speach-enhancement}. Researchers have explored various optimization strategies for Processing Engine(\textbf{\PE}) design across different abstraction levels to enhance resource efficiency and throughput, though this is often achieved at the expense of accuracy~\cite{vachhani2009efficient-CORDIC-Arch-low-area-high-throughput}.

Therefore, designing circuits involves balancing area, power, latency, throughput, and accuracy. This study promotes the implementation of the Iterative and recursive \cordic algorithm, facilitating \mac computation with minimal hardware resources and lower power consumption. The accuracy of recursive \cordic-based \mac and \AF is determined by the number of iterations; increased iterations enhance accuracy and primarily influence throughput~\cite{sumiran-SoftMax}. The author has incorporated the adaptable \cordic architecture~\cite{Flex-PE} to enable multifunctional capabilities and maximize the use of \cordic. The proposed performance enhancement block ultimately provides increased flexibility and control over hardware design, benefiting from a Systolic array~\cite{SANoC}, which provides high throughput. In \dnn and Transformers, performance metrics are limited by the \mac operation, a primary computational aspect that consumes significant resources. Typically, advancing architectural efficiency requires numerous \macs— VGG-16 involves 15.5 billion \macs, AlexNet utilizes 724 million \macs, ResNet-50 necessitates 3.9 billion \macs, while Transformers like GPT-3 and GPT-4 demand 175 billion and up to 1.76 trillion parameters, respectively, as noted in Table \ref{tab: model parameters}. Collectively, these points highlight the need for a reconfigurable design capable of supporting Transformers, \dnns, and \RNNs/LSTMs, providing control at the level of individual Reconfigurable Processing Engines (RPE). A modifiable RPE block enhances hardware flexibility for diverse kernel sizes and \AFs, enabling the accelerator to handle most model computations and reducing reliance on the CPU.

\begin{table}[]
\caption{State-of-the-art AI Models and their corresponding \mac Units and parameters with the necessity for Hardware Optimization}
\label{tab: model parameters}
\begin{tabular}{c|c|c|c|c|c}
\hline
\textbf{Model} & \textbf{Dataset} & \textbf{Parameters} & \begin{tabular}[c]{@{}c@{}}\textbf{MAC count} \\ (\textbf{Billions})\end{tabular} & \begin{tabular}[c]{@{}c@{}}\textbf{Sparsity}\\ (\%)\end{tabular} & \begin{tabular}[c]{@{}c@{}}\textbf{Physical Overhead} \\ (\textbf{Area-Power-Delay})\end{tabular} \\ \hline
VGG-16 & ImageNet & 140 M & 15.5 & 80 & Medium \\ \hline
ResNet-50 & ImageNet & 28 M & 3.7 & 65 & High \\ \hline
Inception-v3 & ImageNet & 24.6 M & 5.3 & 60 & High \\ \hline
MobileNet-v1 & ImageNet & 4.2 M & 0.6 & 35 & High \\ \hline
BERT-base & SST-2 & 105 M & 21.2 & 55 & Medium \\ \hline
MobileBERT & SST-2 & 26 M & 4.3 & 65 & High \\ \hline
GPT-2 & SST-2 & 118 M & 42 & 35 & Medium \\ \hline
GPT-3 & \begin{tabular}[c]{@{}c@{}}Internet-scaled\\ data\end{tabular} & 175 B & 6.7 & 75 & High \\ \hline
GPT-4 & \begin{tabular}[c]{@{}c@{}}Internet-scaled\\ data\end{tabular} & 1760 B & \textgreater{}1000 & 50 & High \\ \hline
\end{tabular}
\end{table}

The noteworthy contributions of this work are outlined below:  

\begin{itemize}  
\item{\textbf{Reconfigurable Processing Engine (RPE):}\\ 
This work introduces a resource-efficient RPE that employs a Pareto-optimal \cordic(5+2) framework characterized by minimized area overhead, facilitating key AI operations such as \mac, Sigmoid, $\tanh$, \softmax, \gelu, \relu, etc. It dynamically shifts between iterative latency and pipelined throughput, optimizing hardware resource usage in response to AI task demands.}  

\item{\textbf{Systolic \cordic engine for Reconfigurability and Enhanced Throughput (SYCore):}\\ 
The introduced SYCore presents a systolic design incorporating the proposed RPE, delivering high throughput and scalable runtime-configurable dense and sparse matrix multiplications. It functions as a fundamental component for AI tasks, including \dnns, Transformers, and \RNNs/LSTMs.}  

\item{\textbf{Configurable and Adaptive Execution Scheduler for Advanced Resource Allocation (CAESAR):}\\ 
The newly developed CAESAR control unit adopts an adaptive tiling and scheduling approach that dynamically leverages quantization and pruning co-design benefits. This ensures efficient resource utilization and adaptable workload handling in edge-AI accelerators. Several benchmark analyses are provided for detailed result assessment.}  

\item{\textbf{Empirical Analysis for Performance Enhancement:}\\ 
We analyzed the minimal pipelined \cordic stage requirement within RPE to augment SYCore's performance. The custom bitwise Pareto analysis of the \cordic stages underscores their influence on error metrics, accuracy, and hardware characteristics. The outcomes reveal improved hardware efficiency with resource optimization and minimal latency impact, supported by extensive comparisons with similar FPGA and ASIC accelerator benchmarks.}  
\end{itemize}

\subsection{Organization of the paper}
    
The remainder of this paper is organized as follows:

\begin{itemize}

    \item \textbf{Section II: Reconfigurable Processing Engine (RPE) as a Neuron for AI Architectures} \\
This section introduces the proposed \cordic-based RPE, detailing its architectural design, algorithm, and adaptive reconfigurability—the detailed Pareto analysis with design trade-offs associated with error metrics and hardware resources with iterative and pipelined configuration. The sections detail the 5+2 stage \cordic neuron based on RPE, with data and control signal management, to enable precision scaling, computational accuracy, and efficient resource utilization.

    \item \textbf{Section III: SYCore Array and CAESAR Control Engine} \\
This section discusses the parameterized \textit{Systolic \cordic Engine for Reconfigurability and Enhanced Throughput (SYCore)}, controlled with \textit{Configurable and Adaptive Execution Scheduler for Advanced Resource Allocation (CAESAR)}. The primary focus is dynamic workload scheduling with tiling, memory protocols, and kernel mapping within optimized resources.

\item \textbf{Section IV: Evaluation Setup and Comprehensive Analysis} \\
This section outlines the experimental setup and validation framework for different datasets, AI models, and hardware evaluation for FPGA and ASIC platforms with performance. A comprehensive analysis of the proposed architecture with State-of-the-art(\textbf{\sota}) solutions has been conducted. 
    
\item \textbf{Section V: Conclusion and Future Work} \\
Finally, this section summarizes the paper's contributions, discusses potential areas for improvement, and outlines directions for future research.

\end{itemize}

\section{Proposed Neuron Architecture (RPE)}
\label{sec_proposed_solution}
For AI systems that are extensively used, the design must prioritize both energy efficiency and a certain level of flexibility. This would allow handling various tasks using a uniform core, eliminating the need for distinct accelerators for different functions. We have designated the proposed core as the Systolic \cordic engine, named for its Reconfigurability and Enhanced throughput (SYCore). A review of existing literature reveals a crucial gap in hardware design: a lack of flexibility. Many existing architectures do not effectively offer \PE control to users, often only limiting to features like power or clock gating and enabling signals. Nevertheless, through a sophisticated design, better \PE control can be achieved via \cordic blocks. These blocks are equipped to handle computations required by a plethora of AI processes today, including but not limited to \dnns, \RNNs, and Transformers, hence they are suitable for a wide range of AI tasks. The challenge in creating a \cordic-based Reconfigurable Processing Element (RPE) lies in managing the balance between latency and resource consumption, especially during repetitive calculations for \mac operations and \AFs. Linear \cordic stages underpin \mac computation and necessitate multiple iterations for precise results. Although more iterations enhance precision, they concurrently increase latency. Mitigating latency either through fewer iterations or parallel processing escalates resource consumption due to the need for extra hardware, like shifters, adders, and LUTs.

\begin{figure}[h]
    \centering
    \begin{minipage}{0.49\textwidth}
        \centering
        \includegraphics[width=\linewidth,height =50mm]{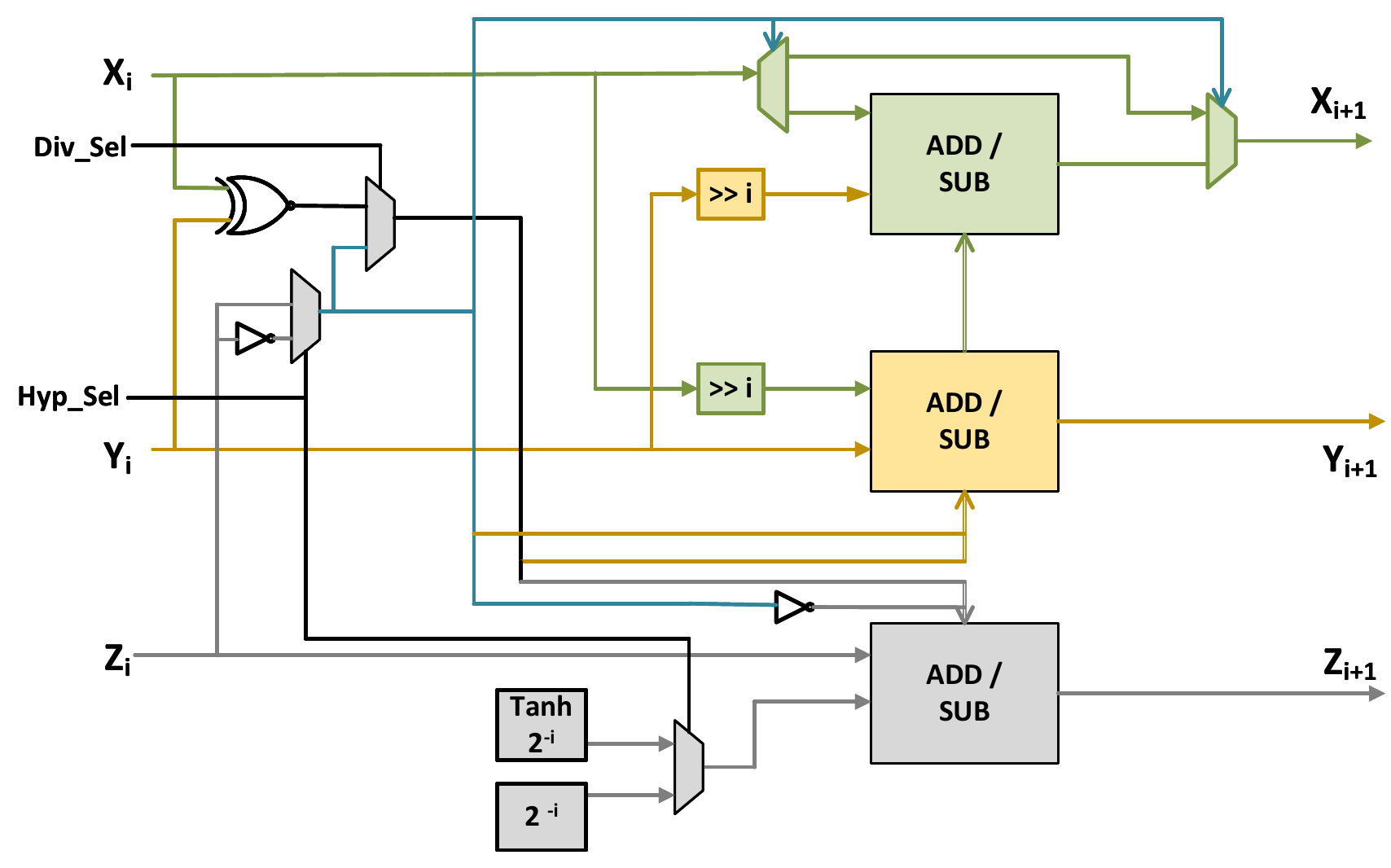}
        
        \subfigure[]{\label{fig:Recon_block}}
    \end{minipage}
    \hfill
    \begin{minipage}{0.49\textwidth}
        \begin{minipage}{\textwidth}
            \centering
            \includegraphics[width=\linewidth,height =25mm]{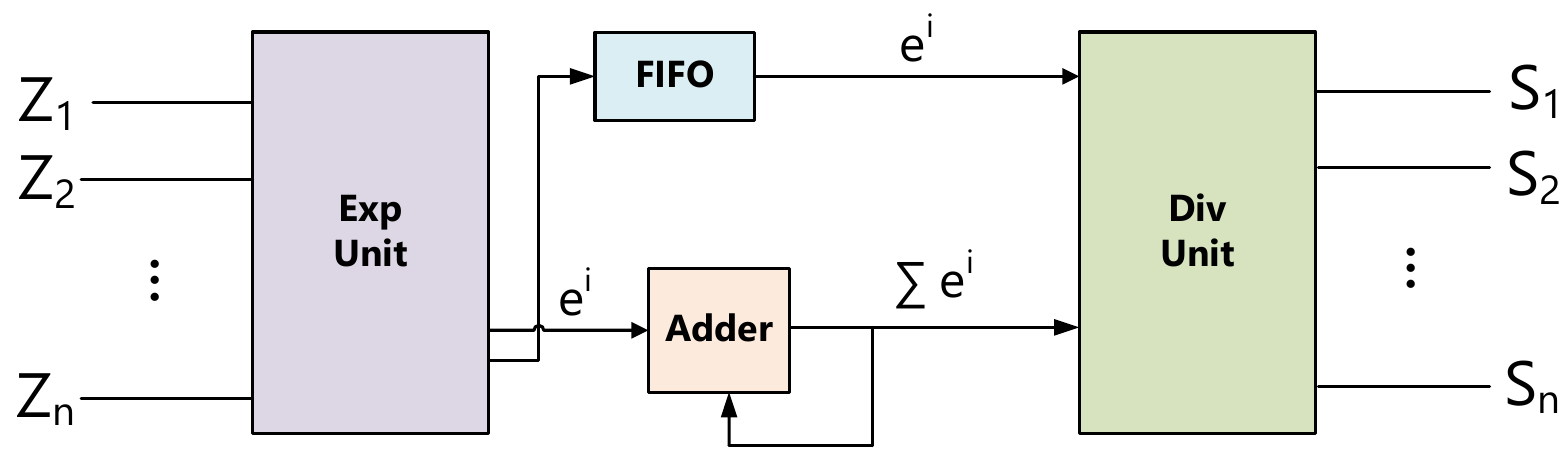}
            
            \subfigure[]{\label{fig:SoftMax}}
        \end{minipage}
        \vfill
        \begin{minipage}{\textwidth}
            \centering
            \includegraphics[width=\linewidth,height =25mm]{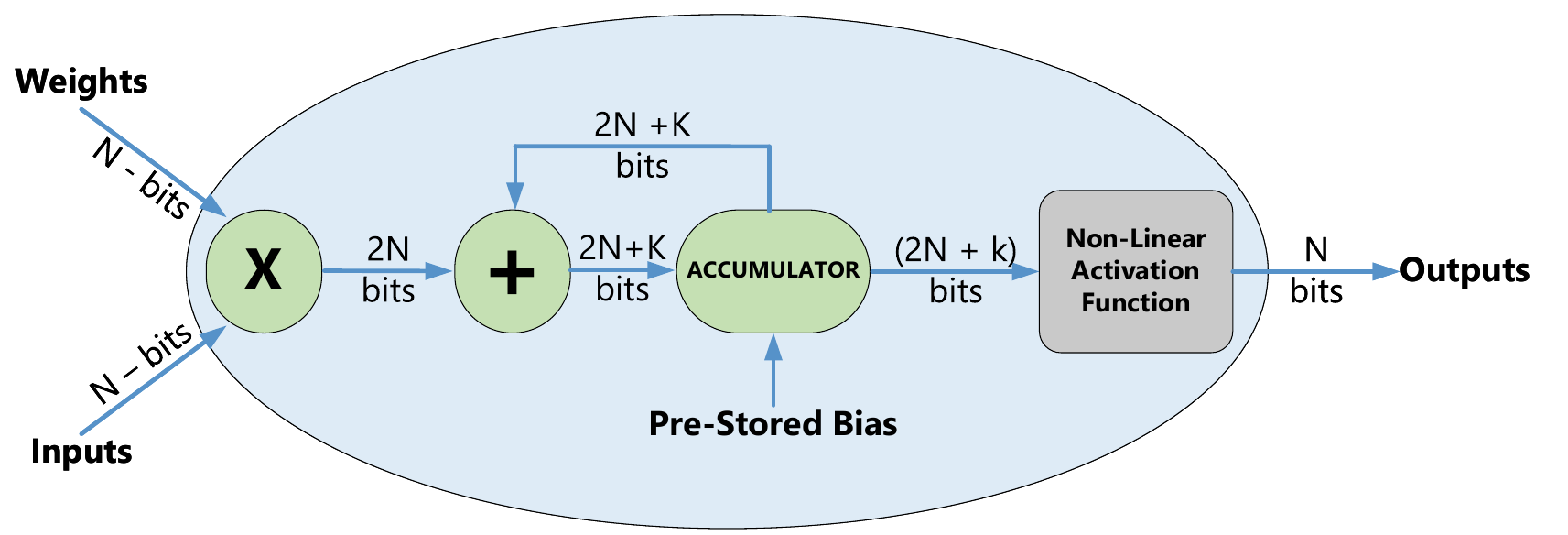}
           
            \subfigure[]{ \label{fig:conventional_PE}}
        \end{minipage}
    \end{minipage}
    
    \caption{a) CORDIC Stage-1 fundamental element b) Reconfigurable Activation function using RPE c) Conventional Processing Engine }
    \vspace{-5mm}
\end{figure}

To effectively compute hyperbolic parameters for various components, such as \AFs (AF) like the sigmoid and $\tanh$, the hyperbolic \cordic stage is typically preferred. This stage necessitates an additional iterative element in its design. Activation functions often rely on pre-stored specific constants and angles, resulting in greater computational demands compared to linear stages. While dedicated hardware for this stage can enhance performance, it also increases resource usage. In contrast, utilizing a shared-stage design can minimize area but may introduce bottlenecks and increase latency. The division stage for \AFs presents similar challenges, with iterative division methods being slower by nature. Employing existing \cordic modules for division conserves resources but diminishes processing speed, whereas specialized division hardware boosts performance at the cost of increased area use. Pipelined hardware can address latency by facilitating parallel execution over iterative processing, though this requires additional registers per iteration. Iterative designs save on hardware by employing the same resources for multiple tasks, which is advantageous for resource-constrained edge devices~\cite{edgezhou}, yet they increase latency, complicating real-time application performance. On the other hand, parallel execution uses duplicate \cordic stages for simultaneous processing, significantly cutting down latency but greatly increasing area and power demands. Precision requirements also remain a pivotal consideration.

Achieving a balance among these trade-offs depends on the specific application. Systems that require low latency, like those in robotics and healthcare, benefit from parallel task execution and reduced iteration counts to meet real-time demands. On the other hand, resource-constrained edge devices focus on iterative approaches that utilize hardware reuse, approximate computations, and scheduling that considers sparsity to conserve resources. A hybrid strategy that combines iterative, parallel, and pipelined methods can provide a balanced architecture, offering flexibility for various workloads. By dynamically optimizing iteration counts and adopting adaptive \cordic designs, it is possible to adjust this balance to accommodate specific or mixed precision requirements within the device's performance limitations.

\subsection{CORDIC algorithm}
\label{subsec_CORDIC algorithm}

\begin{table*}[!t]
\caption{Detailed CORDIC equations: general, linear and hyperbolic }
    \label{cordic_eq}
\resizebox{0.8\textwidth}{!}{
\begin{tabular}{lll}
 General CORDIC & Linear CORDIC & Hyperbolic CORDIC \\

    $\begin{aligned}
   x_{i+1} &= x_{i} - m \delta_{i} y_{i} 2^{-i} \\
   y_{i+1} &= y_{i} + \delta_{i} x_{i} 2^{-i} \\
   z_{i+1} &= z_{i} - \delta_{i} E_{i}
    \end{aligned}$
    \label{cordic_equations}
 &

    $\begin{aligned}
    x_{i+1} &= x_{i} - \delta_{i} y_{i} 2^{-i} \\
    y_{i+1} &= y_{i} + \delta_{i} x_{i} 2^{-i} \\
    z_{i+1} &= z_{i} - \delta_{i} E_{i}
    \end{aligned}$
    \label{equation:cordic_equations_linear}
 &

    $\begin{aligned}
    x_{i+1} &= x_{i} + \delta_{i} y_{i} 2^{-i} \\
    y_{i+1} &= y_{i} + \delta_{i} x_{i} 2^{-i} \\
    z_{i+1} &= z_{i} - \delta_{i} E_{i}
    \end{aligned}$
    \label{equation:cordic_equations_hyperbolic}
\end{tabular}}

\end{table*}

Coordinate Rotational Digital Computer (CORDIC) is an algorithm that mimics rotation for iterative calculations of various linear and nonlinear functions. It is engineered to enhance hardware efficiency, factoring in resource constraints and throughput needs. The quantity of \cordic stages required depends on the fractional bit precision in the adaptive fixed-point format, which encompasses both fractional and integer bit widths. The detailed CORDIC algorithm\cite{raut2023empirical-ACM, Flex-PE} is common knowledge and thus, avoided the explanation here. However, we have added appendix A. A comprehensive explanation is provided in the following subsections using Pareto analysis to identify the optimal number of stages, ensuring robustness against errors despite approximations. The \AF part of the proposed RPE is illustrated in \figref{fig:SoftMax}.

\begin{subequations}
    \label{eq:main5} 
    \begin{align}
        e^{-\text{AFin}} &= \cosh(\text{AFin}) - \sinh(\text{AFin}) \label{eq:5a} \\
        \tanh(\text{AFin}) &= \frac{\sinh(\text{AFin})}{\cosh(\text{AFin})} \label{eq:5b} \\
        \text{sigmoid}(\text{AFin}) &= \frac{1.0}{1.0 + e^{-\text{AFin}}} \label{eq:5c}
    \end{align}
\end{subequations}

\subsubsection{CORDIC-Based Reconfigurable Processing Element (RPE)}:
The \cordic-based RPE serves as the core of the processing engine, enabling efficient iterative calculations for both matrix operations and \AFs. Each RPE is equipped with multiple pipelined linear \cordic stages, which are followed by stages for hyperbolic and division computations. This setup allows for iterative \mac operations along with non-linear \AFs such as sigmoid, $\tanh$, and \softmax. This modular structure supports runtime reconfigurability, making it adaptable to various workloads with differing computational requirements.

\begin{figure*}
    \centering
    \includegraphics[width=0.8\linewidth]{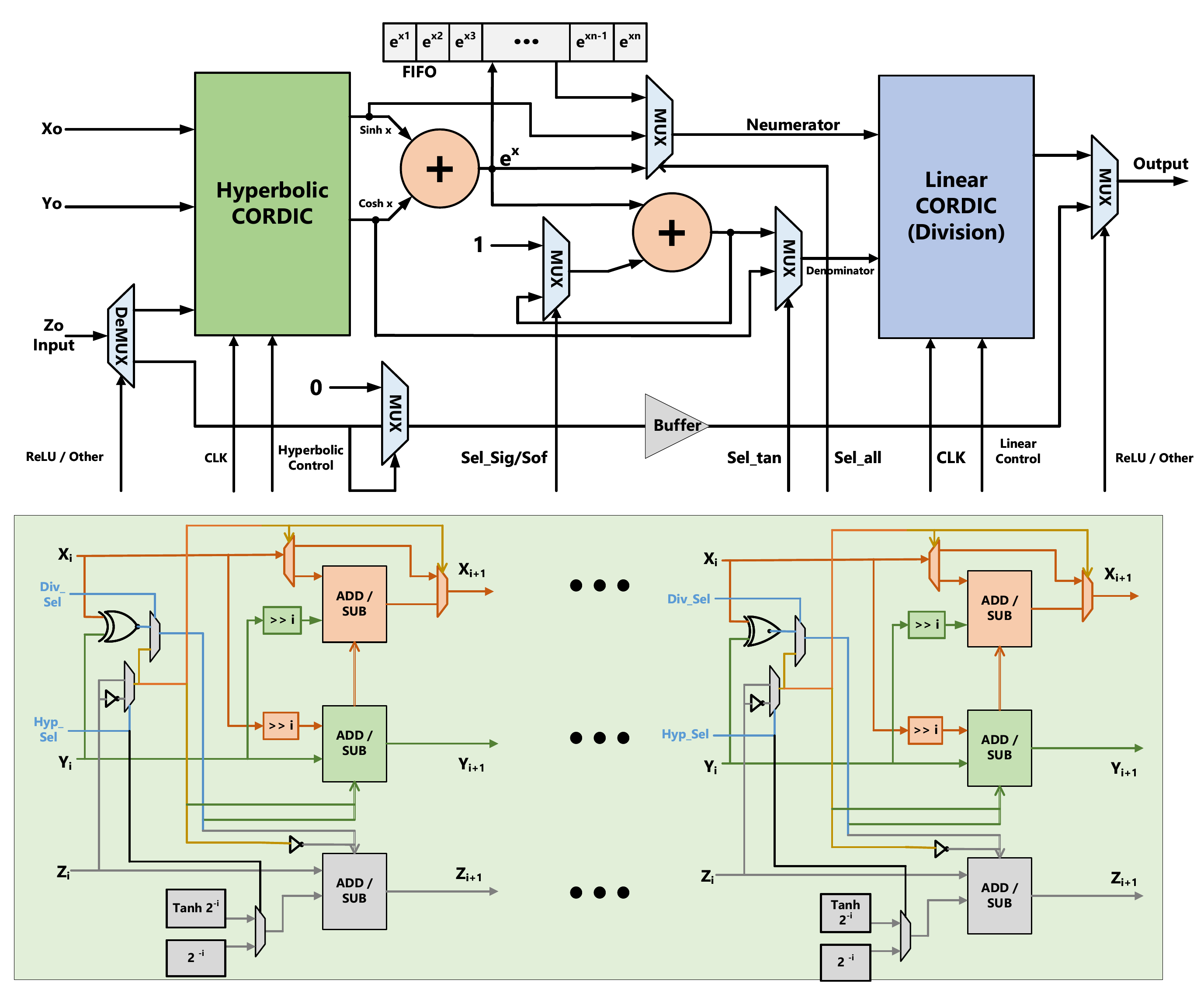}
    \caption{Reconfigurable Activation function using RPE}
    \label{fig:reconfigurable_AF_RPE}
\end{figure*}

\subsubsection{Modes}:
\label{subsec_Modes}
CORDIC, a coordinate rotation algorithm, adapts its operational mode according to the input data type. Initially, \cordic was predominantly used for computing non-linear functions, such as tanh and sigmoid, which are integral to many deep neural networks (DNNs). The \cordic algorithm encompasses three primary modes: hyperbolic, linear, and circular. These modes enable a broad spectrum of operations, from simple multiplication and division to intricate calculations like logarithmic and hyperbolic operations, such as $\sinh$ and $\cosh$, performed within the rotational \cordic structure. Additionally, \cordic offers a Vector Mode for computing the inverse functions of these operations. The foundational \cordic equations represented in Table \ref{cordic_eq} assign the mode values of 0, -1, and 1, which correspond to circular, hyperbolic, and linear modes, respectively. These mode indicators are stored in $m$ to identify the mode during the transition process. The inputs and outputs adjust dynamically with the selection of multiplication, division, or hyperbolic operations. In the Linear Mode, used for multiplication and division, $2^{-i}$ is calculated alongside variable inputs for different \mac operations. The $X$ signal is designated for inputs, the $Z$ signal for weights, and the $Y$ signal can be utilized for additions, as described in \eqref{equation:cordic_MAC} below:
\begin{equation}
    \mac(input=X_0, weight=Z_0, bias=Y_0 )=Y_0+X_0*Z_0
    \label{equation:cordic_MAC}
\end{equation}

For Hyperbolic Mode, \tblref{cordic_eq} illustrates the process where the input, provided through the $Z$ input, results in the computation of $\sinh$ and $\cosh$ at $X$ and $Y$ outputs. Depending on the sign of $Z$, the variable $d_i$ determines whether additions or subtractions are performed on the values of $X$, $Y$, and $Z$.

\begin{equation}
\begin{aligned}
 \text{SoftMax}(a) = \frac{e^a}{\sum_{i=0}^{n} e^i}  \quad  a\in (0,n)
\end{aligned}
\label{equation: softmax}
\end{equation}

\subsubsection{Pareto Analysis}
\label{subsubsec_Pareto Analysis}
\begin{figure}[h]
    \centering
    \begin{minipage}{0.33\textwidth}
        \centering
        \includegraphics[width=\linewidth]{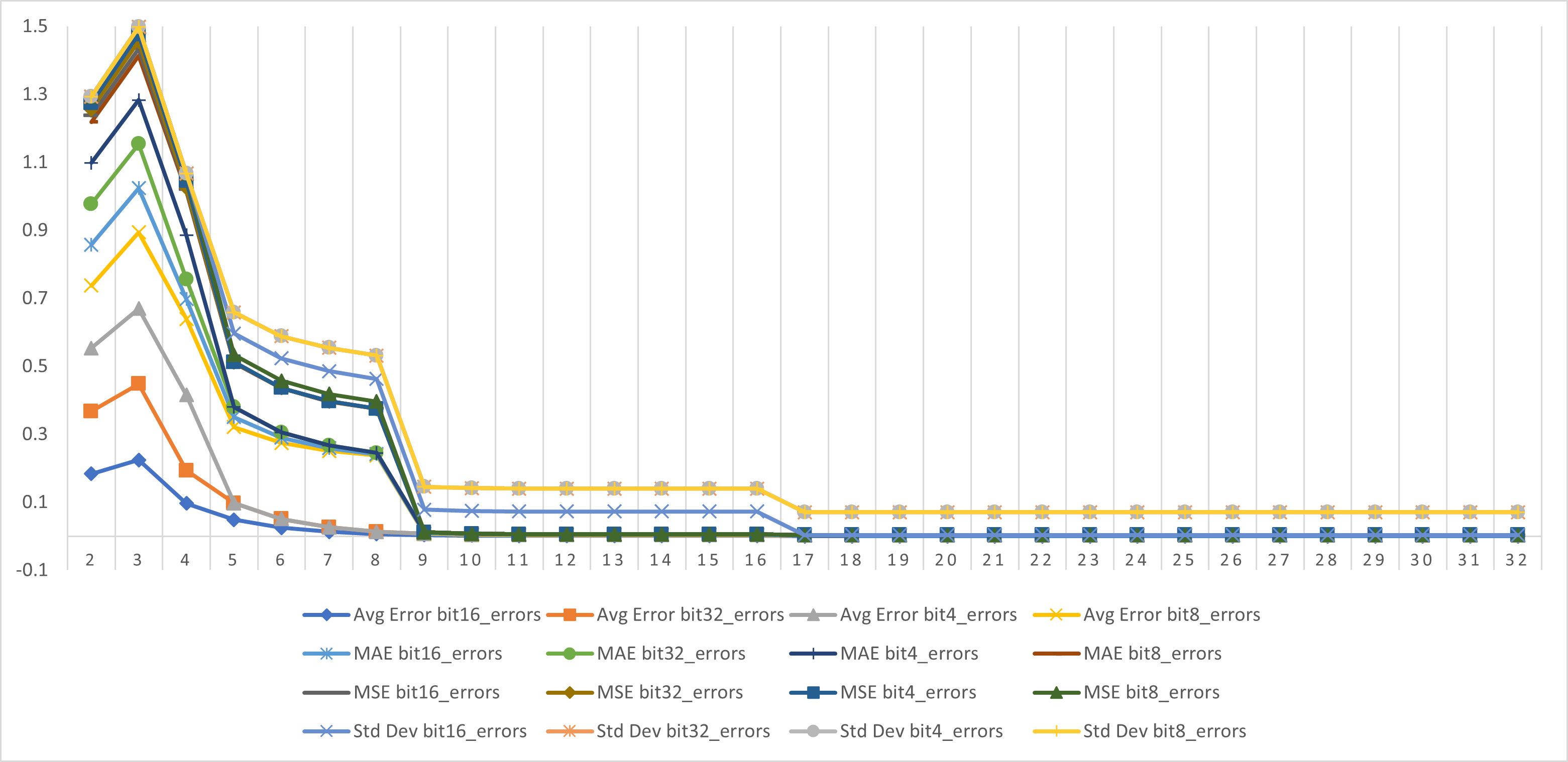}
        \caption{sigmoid error plot}
        \label{fig:sigmoid_error_plot1}
    \end{minipage}
    \hfill
    \begin{minipage}{0.33\textwidth}
        \centering
        \includegraphics[width=\linewidth]{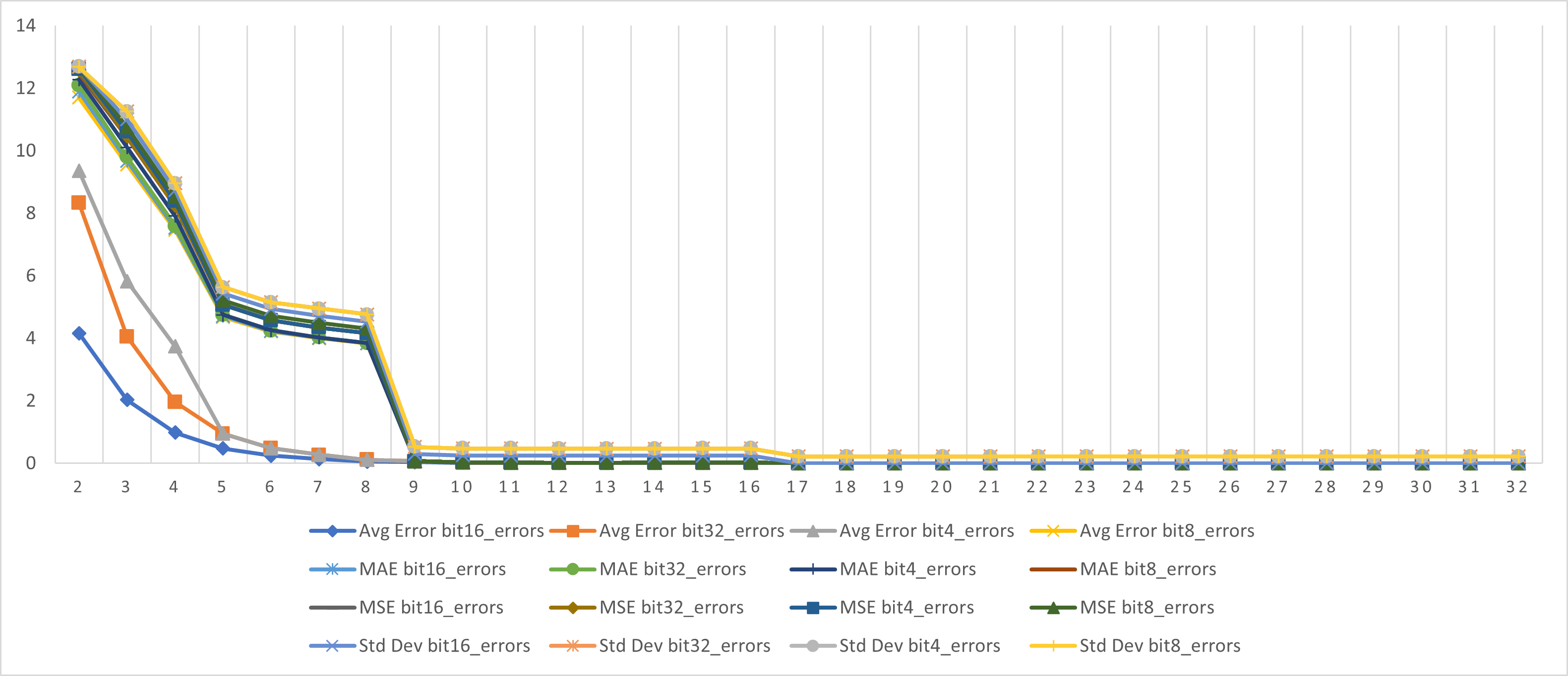}
        \caption{$\tanh$ error plot}
        \label{fig:tanh_error_plot}
    \end{minipage}
    \hfill
    \begin{minipage}{0.33\textwidth}
        \centering
        \includegraphics[width=\linewidth]{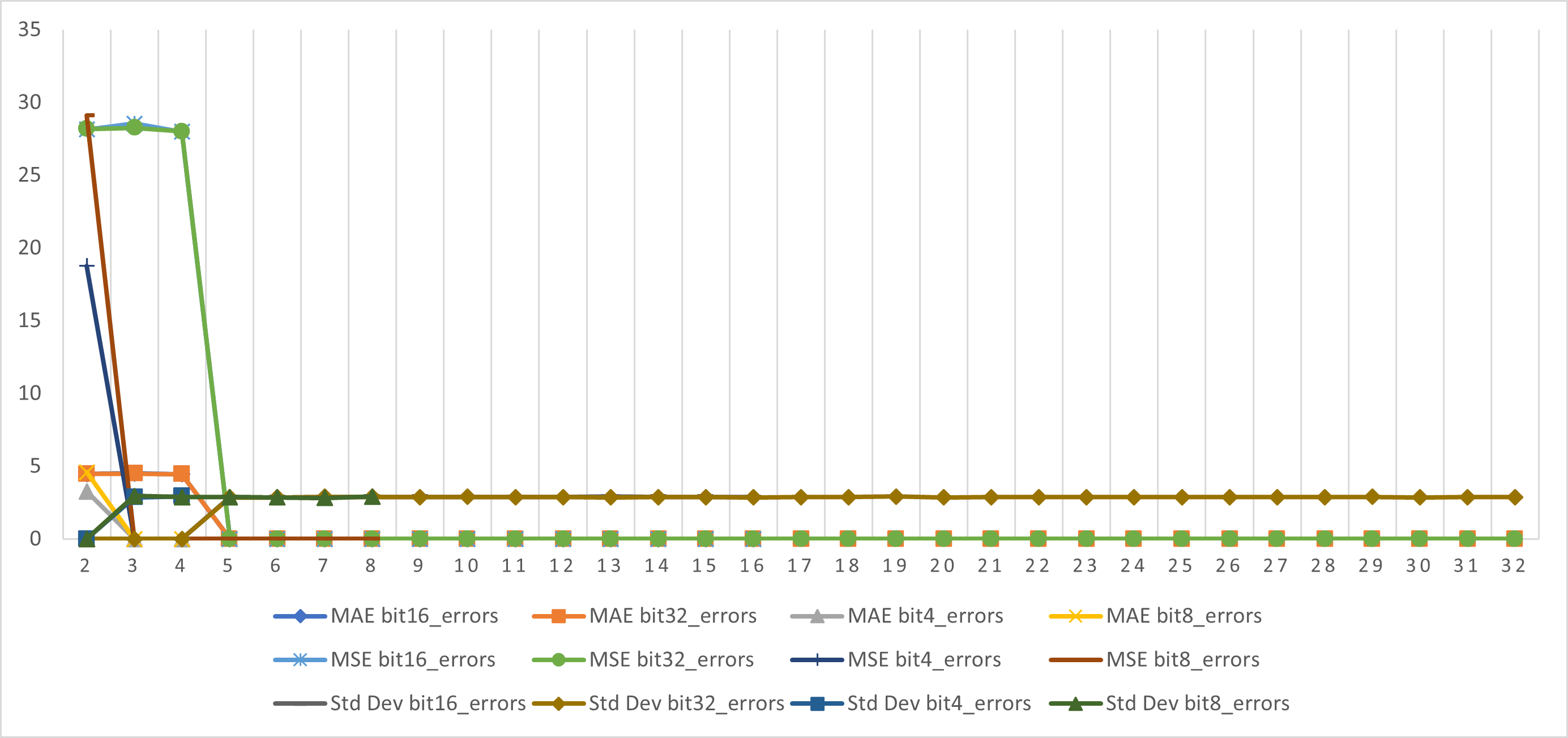}
        \caption{SoftMax error plot}
        \label{fig:SoftMax_error_plot}
    \end{minipage}
    
\end{figure}

CORDIC is a pseudo-rotational framework that induces computational errors since it estimates values through an iterative process, with more stages resulting in greater accuracy. However, increasing iteration count also raises hardware latency, thereby decreasing throughput. Historically, a large number of iterations slowed computation, especially for \AFs. The advent of Transformers has amplified the computation demands for \AFs. To determine the optimal number of iterations across all utilized functions, we have simulated \cordic error across various precisions: 8, 16, and 32 bits, for different iteration counts, as illustrated in the Pareto plots: \figref{fig:tanh_error_plot}, \figref{fig:sigmoid_error_plot1}, and \figref{fig:SoftMax_error_plot}. The Pareto analysis suggests that beyond a specific iteration count, error reduction becomes negligible, thus supporting the adoption of a limited number of iterative stages. The evaluation was conducted using Python within Jupyter Notebook, leveraging libraries such as fxpmath, numpy, and others\footnote{The code is available in the GitHub repository \url{https://github.com/OmkarRajeshKokane/CORDIC-Is-All-You-Need}. }




The \cordic algorithm, primarily assessed for iterative calculations in \AFs and trigonometric operations, was evaluated for potential approximations balancing accuracy, resource usage, and energy efficiency. Consequently, Pareto analysis reveals that optimal configurations are contingent upon specific workload needs. This approach leads to a minimal-error design, mitigating accuracy degradation in \cordic computation by cutting unnecessary iterations and integrating Pareto analysis with bit precision considerations. This study highlights the proposed \cordic design's flexibility, allowing for application-specific adjustments to optimize performance, power, and accuracy constraints.  \figref{fig:sigmoid_error_plot1}, \figref{fig:tanh_error_plot}, \figref{fig:SoftMax_error_plot} illustrates the errors across varying bit precisions (4, 8, 16, 32 bits) over iterations, presenting metrics such as average error, mean absolute error (MAE), mean square error (MSE), and standard deviation (STD). The formulas for these errors are provided in the following equations:

\begin{table}[h]
\centering
\resizebox{0.8\textwidth}{!}{%
\begin{tabular}{cc}
\begin{minipage}{0.45\textwidth}
\begin{equation}
    \text{MSE} = \frac{\sum_{i=1}^{n}\left( y_i-x_i\right)^2 }{n} 
    \label{equation:MSE_error}
\end{equation}
\end{minipage}
&
\begin{minipage}{0.45\textwidth}
\begin{equation}
    \text{MAE}= \frac{\sum_{i=1}^{n}|y_i-x_i|}{n}
    \label{equation:MAE_error}
\end{equation}
\end{minipage}
\\[10pt]

\begin{minipage}{0.45\textwidth}
\begin{equation}
    \text{Avg. \ Error} = \frac{\sum_{i=1}^{n}\frac{|y_i-x_i|}{|x_i|}}{n}
    \label{equation:Avg_error}
\end{equation}
\end{minipage}
&
\begin{minipage}{0.45\textwidth}
\begin{equation}
    \text{STD} = \frac{\sum_{i=1}^{n} [x_i- \text{mean}(y)]^2 }{n-1}
    \label{equation:STD}
\end{equation}
\end{minipage}
\\
\end{tabular}%
}
\end{table}

These equations represent the formula utilized for determining Pareto errors, with \(x\) denoting the anticipated output and \(y\) indicating the resultant output utilizing Fixed-point \cordic design. The error is characterized as the average discrepancy in the values. Larger values correspond to greater error magnitudes, and the same error percentage translates into larger absolute errors for higher values, as compared to those shown in Pareto plots. 

\subsection{Proposed 5+2 RPE Architecture}
\label{sec_Proposed RPE Architecture}
 The Multiply-Accumulate (MAC) unit is the most frequently utilized component in AI accelerators. To improve the \mac's performance with \cordic, we have engineered a \cordic design that executes the \mac operation within the systolic array architecture, as delineated previously in \secref{subsec_Modes}. The substantial delay within \cordic arises from its inherent iterative design. To address this, a pipelined architecture has been introduced to the \mac unit, enhancing throughput. The \mac operates in a five-stage pipeline, thereby decreasing the circuit latency, with five clock cycles required to produce an output. Initially, the multiplication process occupies five clock cycles, with each cycle computing a multiplication. This allows multiplication operations to occur at frequencies of 3\GHz or higher, enabling three billion multiplications per second. Primary operations in convolutional layers, fully connected networks, or transformers include multiplication in 2D/1D convolution and matrix multiplication formats. Additionally, the \AF is implemented using a \cordic iterative technique, prioritizing area efficiency and computational necessity. The versatile design enables \cordic to execute various \AFs such as $\tanh$, Sigmoid, \relu, and the commonly used \softmax. Implementation demands the division phase of \cordic, along with two adders and several Register Array memory units operating in a FIFO configuration to hold \softmax values. Multiple multiplexers and demultiplexers are utilized to select the desired \AF.

\subsubsection{CORDIC: Pipelined vs Iterative analysis}
\label{subsec_CORDIC: Pipelined vs Iterative analysis}
The \cordic (COordinate Rotation DIgital Computer) algorithm is extensively used for the calculation of trigonometric, hyperbolic, and logarithmic functions. Its hardware can be realized through either pipelined or iterative methods, each presenting unique trade-offs. A pipelined approach offers advantages such as reduced delay in throughput, needing five clock cycles initially to produce the first output, and then providing output multiplied for each subsequent clock cycle. The minimized delay in the combinational path enables higher clock frequency operation, thereby enhancing the entire design's frequency capabilities. However, pipelining requires additional flip-flops between stages, which increases power consumption and physical design area. This method can be adapted for different levels of precision with minimal design alteration, primarily modifying data paths rather than control blocks. On the other hand, iterative \cordic designs prioritize area and power efficiency, leveraging pure combinational logic at the expense of higher delay and lower throughput and are less amenable to frequency scaling. Although additional pipelining can enhance pipeline designs, a hybrid approach facilitates the optimal balance of delay, power, and area. In iterative \cordic, storing angle rotation constants for hyperbolic and division stages in memory is necessary, unlike in pipelined designs where exact stage values are pre-determined. Iterative designs also undergo challenges related to control complexity. The proposed design uses iterative stages for hyperbolic and division functions since these operations are less frequent than multiply-accumulate (MAC) operations. Thus, a five-stage pipeline is implemented for \mac calculations, alongside an iterative design for hyperbolic and division stages.

\subsubsection{5+2 stage architecture}
\label{subsec_5+2 stage architecture}
Pipelining the \mac into five stages alongside two iterative stages for hyperbolic and division calculations enhances the throughput in our \cordic design, without significant compromise on accuracy, as confirmed by Pareto-driven analysis. Each pipelined stage features its own $2^{-i}$ values that correspond to their respective iterations, facilitating logical optimization during netlist generation. Further, benefits of the \cordic pipeline are discussed in \secref{subsec_CORDIC: Pipelined vs Iterative analysis}. Moreover, complex Deep Neural Networks (DNNs) necessitate complex \AFs, thus the employment of iterative hyperbolic and division stages enhances the functionality for \AFs such as $\tanh$, sigmoid, and \softmax. There's also an alternative design that omits the hyperbolic and division stages, thereby lowering latency. However, the \softmax function's integration remains essential for transformer-based architectures, enhancing \softmax processing flow. Flex-PE and DA-VINCI~\cite{Flex-PE, Kokane_2024-SAMRT-AF} expands the range of supported \AFs, including $\tanh$, sigmoid, \softmax, \relu, \selu, Swish, and \gelu, thereby enabling compatibility with diverse AI models like \RNN/LSTM, \dnn, and Transformers. The resulting hybrid architecture, while more intricate to manage, demands a sophisticated design and precise control mechanisms, which we will elaborate on in the next section.

\subsection{Details of Data and Control Signals and and State Machine for Neuron Engine}
\label{subsec_Details of Data and Control Signals}

To manage the design, understanding the data flow within the RPE is essential, starting from the data entry into the registers. Once the enable signal reaches the RPE, the initial stage of the \mac, synchronized with the clock, processes the data. It operates on the variables $x$, $y$, and $z$ and relays the outcome to the subsequent register, where $x$ receives the input values, $y$ acquires the bias, and $z$ obtains the weight values. Depending on the sign of $z$, the term $\delta_{i}$ determines whether to execute addition or subtraction. A shifter maintains the connections. The same procedure is repeated in the next cycle, with variation only in the angular constants $E_i$, which have values of $2^{-i}$ based on the iteration count. This pipelined process continues until the fifth iteration, passing data to the subsequent pipelined block with each iteration. Every clock cycle triggers a new input for \mac operation, producing multiplied outputs at each cycle and capitalizing on the pipelined architecture. Upon completing multiplication and accumulation with all kernel elements, the $hyp_select$ signal activates the hyperbolic stage flow, as depicted in \figref{fig:SoftMax}. Following this, the accumulated data proceeds to the Hyperbolic design for five clock cycles using the existing \cordic hyperbolic stage, controlled by a counter to route the data to the next stage. The division phase of the \cordic design commences with the $Div_select$ signal and a similar counter to issue the output completion signal. Various signals are directed to the multiplexers and demultiplexers. To choose the \AF, the $Relu/other$ signal decides whether to employ \relu, while $sel\_sig/sof$ selects between sigmoid and \softmax. The $sel\_tan$ signal determines whether the design operates with hyperbolic tangent, all affecting the denominator. Conversely, a three-input multiplexer sets the numerator, selecting among the three AFs: $\tanh$, sigmoid, and \softmax via $sel\_all$. \softmax computation diverges slightly: inputs stored in the FIFO memory are summed during storage. These stored inputs are then fed to the Division \cordic block to yield the \softmax values, concluding the RPE process with the $RPE\_done$ signal.

A suitable strategy for structuring this control flow is through the use of a \fsm. To devise an \fsm controller for the RPE, certain parameters are initially necessary, such as kernel size and the specific operation to be executed within \AF, which are critical in the setup phase. Starting from the initial stage, the system proceeds to the initiation state, where it accepts data inputs and iterates over five clock cycles, continuously receiving inputs and weights each cycle while keeping the other \AF signals inactive. Once the first multiplication outcome reaches the final pipeline stage, the \fsm progresses to the following phase—namely, the Rigorous \mac operation stage—where inputs for multiplication are continually processed in subsequent iterations until all multiplications are finalized. Upon completing the multiplication phase, the \fsm transitions to the \AF, determined by the selected \AF. In the scenario where $\tanh$ or sigmoid is chosen, the \fsm enters case 1, beginning with the Hyperbolic computation minor stage, followed by the Division stage in its div state. The sequence concludes with the RPdidne signal, returning to the ideal state. For case 2, if the user opts for the \softmax state, all inputs undergo hyperbolic processing in the first stage, being stored in the \softmax FIFO concurrently, after which the process evolves to the division state. Here, the FIFO-stored values are utilized to derive the exponent terms divided by the cumulative output generated during the hyperbolic stage, summing all input exponents. This division is executed iteratively for each FIFO input, taking $n$ times 4 clock cycles, culminating in the completion of case 2 and dispatching the RPE complete signal. For the simplest case, 3, the \fsm selects the \relu flow, bypassing unnecessary iterations of 9 clock cycles in hyperbolic and division stages, thus conserving clock cycles and enhancing throughput. Completing from a singular \AF state of \relu, the \fsm reverts to the IDLE state. When integrated into larger applications, such designs facilitate executing a broad array of tasks on the same hardware, encompassing LLMs, computer vision tasks, and more. However, to meet throughput demands, a lone RPE block is insufficient. The architectural framework is crucial in optimizing RPE's advantages, ensuring the design's applicability and maximizing efficiency in addressing these requirements.

\subsection{Optimizations for Next-Generation Workloads: DA-VINCI}
The sophisticated \AF architecture of the DA-VINCI core enables seamless incorporation of \AFs with minimal computational overhead. These functions can be dynamically configured at runtime to switch among Swish, \softmax, \selu, \gelu, Sigmoid, $\tanh$, and \relu, unified within a single design utilizing \cordic. Prior studies~\cite{Kokane_2024-SAMRT-AF} highlight the efficacy of these functions with a Quality of Results (QoR) reaching 98.5\%. 
The unified CORDIC algorithm~\cite{RECON-CORDICNeuron, Unified-CORDIC} has been preferred for reconfigurable hardware for circular, linear, and hyperbolic operations combined with rotational and vector modes. The fundamental CORDIC hardware uses simple design elements such as Add/Sub, MUX, LBS, and memory blocks. 
The HOAA-enabled CORDIC architecture~\cite{HOAA} provides 21\% area savings and up to 33\% lower power consumption, effectively compensating for the area overhead introduced with DA-VINCI core's reconfigurability. Thus, the proposed DA-VINCI core almost matches the resource utilization for individual AF implementations.  
The detailed methodology for CORDIC-based activation function hardware has been explored for Swish~\cite{ReAFM-NN}, SoftMax~\cite{SoftAct-Trans}, SELU~\cite{Designspaceexploration-AF}, GELU~\cite{ReAFM-NN}, Sigmoid~\cite{CORDICAF-LSTM}, Tanh~\cite{CORDICAF-RNN}, and ReLU. DA-VINCI programmable core is built on linear and hyperbolic CORDIC block, with sel\_af signal, program particular AF datapath and sequence of operations. sinh and Cosh are computed with HR mode in Swish, SoftMax, SELU, GELU, Sigmoid and Tanh, marking 86\% reuse factor, while division operation with LV mode in Swish, SoftMax, GELU, Sigmoid, and Tanh marking 72\% reuse factor. Additional buffer in case of ReLU, FIFO in softmax and a couple of multiplication units for GELU, and fitting additional SELU and Swish contribute to extra overhead, which identifies as hardware vs reconfigurability tradeoff. 

DA-VINCI surpasses current designs in resource efficiency, achieving notable reductions: up to 4.5\X in LUT usage, 3.2\X in flip-flop (FF) usage, 7.8\X in critical path delay, and 14.3\X in power consumption. Both FPGA and ASIC analyses demonstrate these advantages, with ASIC showing up to 16.2\X area reduction, 7.8\X delay reduction, and 14.3\X power reduction across different technology nodes and bit precisions, as previously discussed in our works. Integrating the DA-VINCI Design with the RPE block enhances its capability to handle diverse workloads, making it feasible to implement every non-pooling layer using a consistent design due to its support for \AFs like \softmax and \gelu in Transformers and Swish and \selu in \RNNs/LSTMs, without excluding \relu, Sigmoid, and $\tanh$. This broadens the design's acceleration potential across these functions. The design incorporates additional multipliers, which can also be built using \cordic, along with an HOAA adder to address overestimated addition~\cite{HOAA}, and several multiplexers for \AF selection. Initial multipliers in the \AF handle scaling of the matrix multiplication (\mac) outputs, which involves scaling non-linear output values. Such operations can extend across layer transitions, as observed in transformer models that contain normalization blocks. The hyperbolic stage supplies exponential and hyperbolic outputs via adders, while division tasks rely on the division capabilities of the \cordic stage. This allows most AI-driven operations to be efficiently executed through a fundamental \cordic block, from conventional \mac procedures to \AFs like \selu and Swish. Additionally, \cordic is proficient in executing logarithmic and square-root functions among others, indicating that \cordic is the sole fundamental block necessary for all workloads, validating the adage \cordic IS ALL YOU NEED.

\begin{figure}
    \centering
    \includegraphics[width=\linewidth]{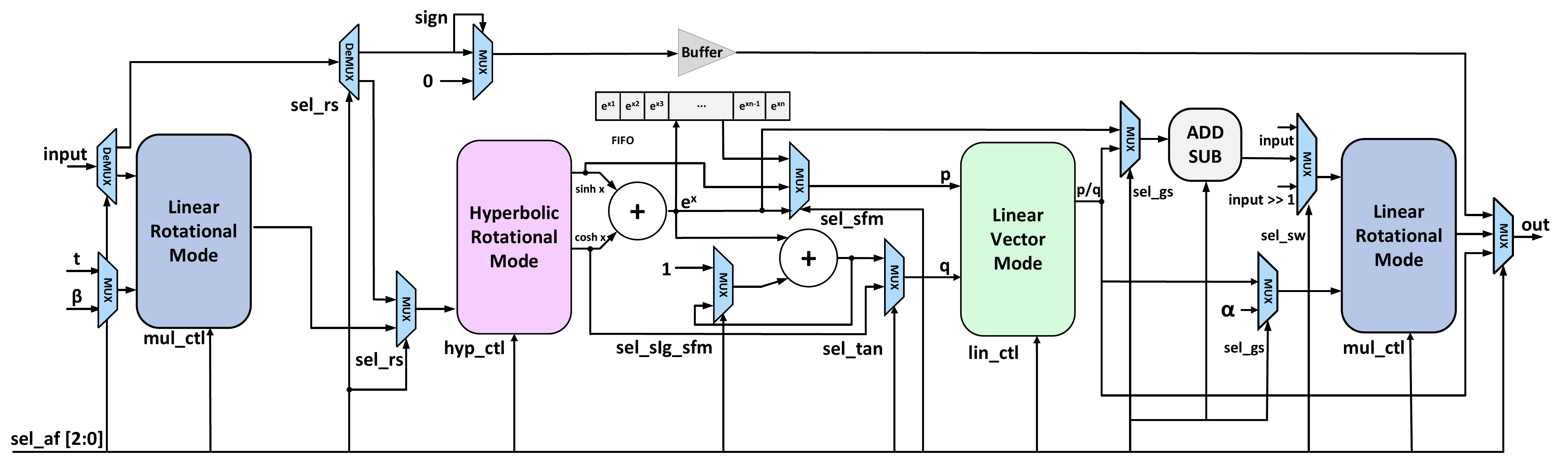}
    \caption{Dynamically-configurable activation function supporting GeLU, SeLU, Swish, ReLU, Tanh, Sigmoid and Softmax}
    \label{fig:enter-label}
\end{figure}

\section{Parameterized and Modular Systolic Accelerator Architecture}
\label{sec_Parameterized and Modular Systolic Accelerator Architecture}
The comprehensive analysis of a complete FPGA design flow contributes significantly to an in-depth understanding of the entire process. Starting from the initial software code, the model is developed in TensorFlow/Torch and executed on the proposed RPE using a Systolic Array architecture. In \secref{sec_Proposed RPE Architecture}, we explored the RPE design and its advantages. We now aim to examine the Systolic array's design and evaluate its benefits across different types of flows, comparing it with other accelerator architectural designs. This includes architectures like Layer-reuse, 1D structures, NAS-based designs for task-specific applications, parallel architectures, and sparsity-based designs such as RAMAN~ \cite{RAMAN-IoTJ24}. Additional research in architecture has spawned designs such as Eyeriss~\cite{Eyeriss}, EIE~\cite{EIE}, Feather~\cite{tong2024feather-tk}, Aurora, and more. The primary motivations behind various architectures are to optimize the design for particular processing elements and to harness benefits such as sparsity, pruning, SIMD, throughput demands, and other specific constraints.

Analyzing various architectures individually, starting with the Layer reuse architecture reveals the concept of utilizing the same hardware across multiple layers. This approach is predominantly applied in 1-dimensional arrays of processing element designs where, sequentially, only one layer is processed at a time. The 1-D architecture can be adapted for layer multiplexing~\cite{raut2023empirical-ACM} or time multiplexing~\cite{Kokane_2024-LPRE}, each offering inter-layer parallelism but requiring greater hardware resources. Although this results in lower latency, which is suitable for low-latency applications, it introduces complex data and control flows. In contrast, time multiplexing offers intra-layer parallelism by reusing hardware over time. This method is simpler in flow complexity but incurs high latency due to sequential processing. Nonetheless, its hardware efficiency makes it suitable for devices with limited resources. 

Turning our attention to NAS architectures, researchers have proposed multicore devices where each core is specifically optimized for accelerating distinct tasks, such as CNNs, fully connected layers, pooling, transformers, or normalization and scaling. This approach results in straightforward designs but suffers from inefficiencies in power and area. Conversely, parallel architectures utilizing 1-dimensional \PE arrays facilitate concurrent \dnn model execution, thereby enhancing throughput and reducing latency, which is well-suited for multi-tenant workloads similar to NAS. However, this configuration demands substantial resources and power, making it better suited for high-performance computing (HPC) rather than edge deployments. The Systolic array, characterized by its two-dimensional arrangement of processing elements, functions by keeping input fixed at one \PE and transferring weights across the array. This is advantageous for \dnn applications where the input data is frequently reused. In contrast, a weight stationary approach holds weights constant while the input data traverses the array, useful in scenarios with stable weights, thereby conserving energy on data transfers. Systems using such arrays are often referred to as vector systolic arrays. There are also other structures like output stationary and interleaved architectures. In output stationary designs, partial sums remain stationary while the inputs and weights move in an interleaved manner. Each design aligns with specific workloads, whether it be CNNs, fully connected layers, or \RNNs. The limitations of traditional designs and the incorporation complexities of advanced optimization have spurred the development of solutions like Eyeriss, which continue to evolve, seen in subsequent models like EIE and Feather. These integrate various benefits into a more complex yet efficient system, enhancing throughput. Nevertheless, no existing design exhibits sufficient flexibility to perform multiple operations and run diverse AI workloads on a single architecture. We propose a new design that can manage all operation types and execute different AI workloads using the same framework. To grasp the architecture's intricacies, a comprehensive understanding of the entire system is required.

\subsection{Host Interfacing }
\label{sub_sec_host_interfaceing}

Analyzing the entire workflow from an application perspective enhances the comprehension of the data and control flow within any design. Starting with the sensor component, analogous to the human eye, the machine relies on the camera as its sensor. Various camera attributes are essential, such as its resolution, accuracy, and RGB configuration. Additionally, certain prerequisites are necessary, including the drivers for retrieving data from the camera and the software libraries to import the image into the codebase. The image data is subsequently processed within the flow. Camera FPS is also important. The model employed utilizes the standard quantized TensorFlow-\relu framework~\cite{Tensor-Std}. This compilation of information is directed to the RISC-V compiler~\cite{PULP, cheshire}, which translates the data flow and computations into suitable instructions for the required operations. With CASEAR and SYCore, a mix of custom and standard instructions is integrated. RISC-V is a processor based on a reduced instruction set computer architecture, featuring a 5-stage pipeline of fetch, decode, execute, memory access, and write-back. These stages are part of the execution process on RISC-V, including custom instructions.
To comprehend the custom instructions on RISC-V, it is necessary to understand the Instruction Set Architecture (ISA). RISC-V accommodates 32 and 64-bit instruction widths. Each instruction belongs to a specific instruction type. RISC-V includes various predefined instruction types and allows for customized instructions. The IFMA category includes the Base Integer Instruction Set (I), Integer Multiplication and Division Extension (M), Atomic Instructions Extension (A), Single-Precision Floating-Point Extension (F), and Double-Precision Floating-Point Extension (D), among other predefined instructions. Instruction Formats such as R, I, S, B, U, and J facilitate custom operations, with the possibility to introduce new formats using reserved opcode bits. Open-source tools like RISC-V GCC and LLVM compilers offer support for crafting and optimizing custom data instructions.

\begin{figure}[h]
    \centering
    \begin{minipage}{0.475\textwidth}
        \centering
        \includegraphics[width=0.9\linewidth, height = 75mm]{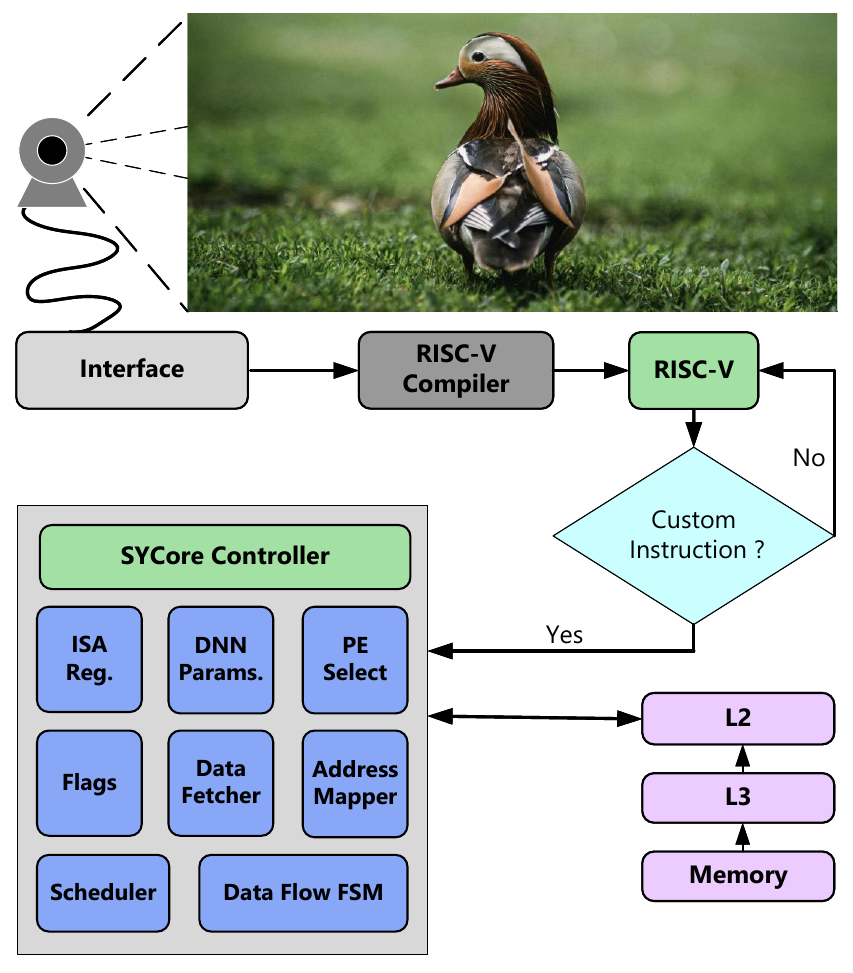}
        \caption{Host Interfacing}
        \label{fig:Host_Interfacing}
    \end{minipage}
    \hfill
    \begin{minipage}{0.475\textwidth}
        \centering
        \includegraphics[width=0.9\linewidth, height = 75mm]{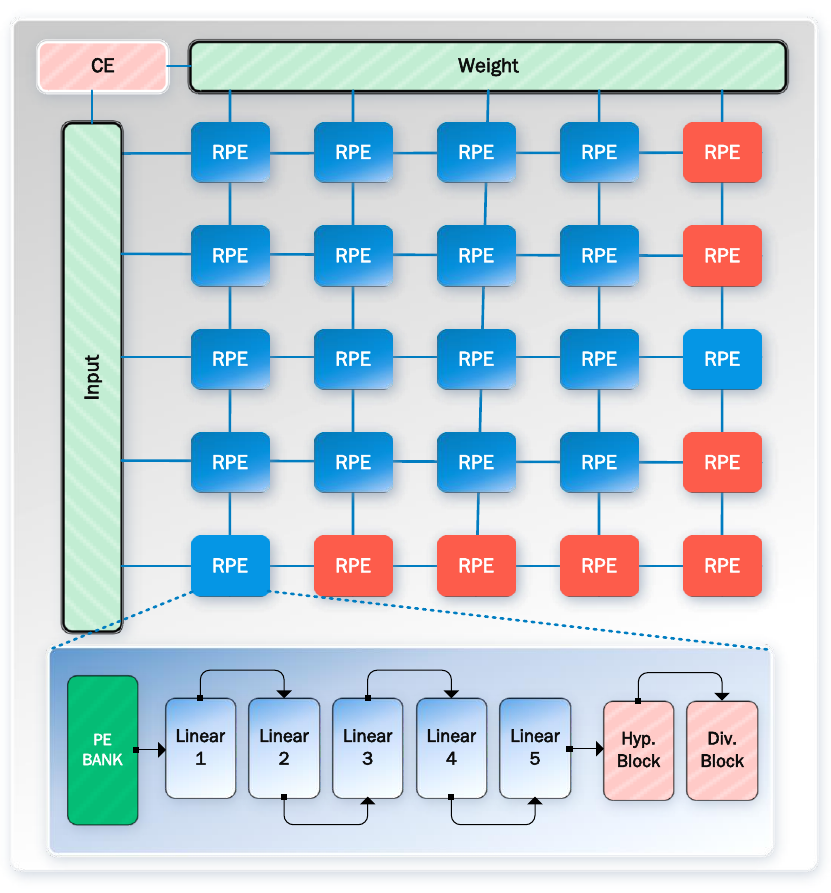}
        \caption{Systolic Array}
        \label{fig:SA-RPE}
    \end{minipage}
    
\end{figure}

Once the Python code is translated into a sequence of instructions, the device determines whether the instruction should proceed to RISC-V pipelining or be directed towards CAESAR. When implementing pooling, the instruction targets the RISC-V path; however, for CNN operations and similar, it is directed to CAESAR, SYCore's control engine. Within CAESAR, the instruction functions as a co-processor, interpreting and leveraging the command to establish data handling procedures, integrating the existing ISA into a register. Subsequently, the necessary \dnn parameters are secured for model implementation by setting flags. The data fetcher begins retrieving data from the shared L2 cache, while the address mapper assigns the retrieved data to appropriate locations. Concurrently, the scheduler determines the RPE requirements, optimizing efficiency in executing the complete routine within the Data-flow \fsm. Upon completion, the output is transferred back to the L2 cache for storage, and a completion signal is issued. This process leads to a reduced workload on the RISC-V, allowing it to perform additional tasks. The workflow is depicted in \figref{fig:Host_Interfacing}.

\subsection{Design and System Architecture Overview}
\label{subsec_{Design and System Architecture Overview}}

An in-depth analysis has been conducted on the comprehensive proposed design, which incorporates SYCore, a stationary systolic array computing output controlled by CAESAR, a sophisticated engine for managing data flow. Beginning with the SYCore is an accelerator developed on a foundational architecture characterized by an output stationary systolic array. This architecture allows the inputs and weights to move continuously for computation processes while maintaining the partial sums in a fixed position. SYCore is comprised of individual RPE blocks configured in a 32\X{32} array and is subdivided into smaller 4\X{4} sub-blocks to facilitate parallel computation and optimize accelerator efficiency. When any sub-block is not in use, all sub-blocks are deactivated to conserve power. Data flow within sub-blocks is enhanced through multiplexers, enabling data transfer to neighbouring sub-blocks. In the output stationary approach, the inputs and weights are transmitted between sub-blocks, minimizing unnecessary data transfer and consequently augmenting the design's energy efficiency.

\begin{table*}[!t]
    \caption{Detailed mapping and scheduling for VGG-16/CIFAR-100 with Proposed RISC-V Enabled SYCore Architecture}
    \label{VGG16-CIFAR100_SA-mapping}
    \resizebox{0.8\textwidth}{!}{%
    \begin{tabular}{c|c|c|c|c|c|c|c}
        \hline
        \textbf{\begin{tabular}[c]{@{}c@{}}VGG-16\\ CIFAR-100\end{tabular}} & \textbf{\begin{tabular}[c]{@{}c@{}}Layer Specifications\\ (KxK)x CinxCout x(HxW)\end{tabular}} & \textbf{\begin{tabular}[c]{@{}c@{}}SYCore\\ (MxN)\end{tabular}} & \textbf{K-MAC Ops} & \textbf{Op. cycles} & \textbf{\begin{tabular}[c]{@{}c@{}}CAESAR-based \\ Utilization (\%)\end{tabular}} & \textbf{\begin{tabular}[c]{@{}c@{}}Execution \\ Time (us)\end{tabular}} & \textbf{\begin{tabular}[c]{@{}c@{}}Power \\ Consumed (mW)\end{tabular}} \\ \hline
        \textbf{C1\_1} & (3x3)x 3x64 x(32x32) & 32x32 & 1728 & 1728 & 100 & 2.65 & 0.63 \\ \hline
        \textbf{C1\_2} & (3x3)x 64x64 x(32x32) & 32x32 & 36864 & 36864 & 100 & 56.7 & 13.55 \\ \hline
        \textbf{P1} & Max-Pool, (2x2), S=2 & - & - & 411 & - & 189 & \multicolumn{1}{l}{} \\ \hline
        \textbf{C2\_1} & (3x3)x 64x128 x(16x16) & 16x16 & 73728 & 18432 & 100 & 28.33 & 6.77 \\ \hline
        \textbf{C2\_2} & (3x3)x 128x128 x(16x16) & 16x16 & 147456 & 36864 & 100 & 56.7 & 13.55 \\ \hline
        \textbf{P2} & Max-Pool, (2x2), S=2 & - & - & 313 & - & 144 & \multicolumn{1}{l}{} \\ \hline
        \textbf{C3\_1} & (3x3)x 128x256 x(8x8) & 8x8 & 294912 & 18432 & 100 & 28.2 & 6.76 \\ \hline
        \textbf{C3\_2} & (3x3)x 256x256 x(8x8) & 8x8 & 589824 & 36864 & 100 & 56.9 & 13.5 \\ \hline
        \textbf{C3\_3} & (3x3)x 256x256 x(8x8) & 8x8 & 589824 & 36864 & 100 & 56.7 & 13.55 \\ \hline
        \textbf{P3} & Max-Pool, (2x2), S=2 & - & - & 515 & - & 237 & \multicolumn{1}{l}{} \\ \hline
        \textbf{C4\_1} & (3x3)x 256x512 x(4x4) & 4x4 & 1179648 & 18432 & 100 & 28.4 & 6.78 \\ \hline
        \textbf{C4\_2} & (3x3)x 512x512 x(4x4) & 4x4 & 2359296 & 36864 & 100 & 56.5 & 13.45 \\ \hline
        \textbf{C4\_3} & (3x3)x 512x512 x(4x4) & 4x4 & 2359296 & 36864 & 100 & 56.66 & 13.55 \\ \hline
        \textbf{P4} & Max-Pool, (2x2), S=2 & - & - & 328 & - & 151 & \multicolumn{1}{l}{} \\ \hline
        \textbf{C5\_1} & (3x3)x 512x512 x(2x2) & 2x2 & 2359296 & 36864 & 33.33 & 56.4 & 13.86 \\ \hline
        \textbf{C5\_2} & (3x3)x 512x512 x(2x2) & 2x2 & 2359296 & 36864 & 33.33 & 56.66 & 13.66 \\ \hline
        \textbf{C5\_3} & (3x3)x 512x512 x(2x2) & 2x2 & 2359296 & 36864 & 33.33 & 56.9 & 13.54 \\ \hline
        \textbf{P5} & Max-Pool, (2x2), S=2 & - & - & 165 & - & 76 & \multicolumn{1}{l}{} \\ \hline
        \textbf{Flatten} & - & - & - & 1 & - & 55.88 & \multicolumn{1}{l}{} \\ \hline
        \textbf{FC6} & 512x4096 & 1024 & 4096 & 2048 & 100 & 3.1 & 0.75 \\ \hline
        \textbf{FC7} & 4096x4096 & 1024 & 4096 & 16384 & 100 & 25 & 6.024 \\ \hline
        \textbf{FC8} & 4096x100 & 256 & 100 & 4096 & 40 & 6.3 & 1.506 \\ \hline
        \textbf{Total} & \multicolumn{1}{l|}{} & \multicolumn{1}{l|}{} & \multicolumn{1}{l|}{} & \multicolumn{1}{l|}{} & \multicolumn{1}{c|}{\textbf{83.75+}} & \textbf{148.6 ms} & \textbf{1.524 W} \\ \hline
    \end{tabular}}
\end{table*}

Within each sub-block, RPE units are organized in a 4\X{4} configuration. Every sub-block contains its unique input and weight buffers, facilitating computations within the same sub-block while transmitting data to neighbouring sub-blocks. Executing these operations concurrently minimizes the design's output latency. By centralizing buffer procedures within sub-blocks, data management is streamlined. This method maintains the scalability of the design by adjusting the number of sub-blocks employed. A design akin to matrix multiplication is effective for diverse workloads, such as Transformers, \RNN/LSTM, and traditional \dnn applications. Each RPE takes 45 clock cycles to produce its first output. To govern this process, we introduce CAESAR, a Configurable and Adaptive Execution Scheduler for Advanced Resource Allocation, which efficiently manages tasks from ISA instruction reading through to execution and processing completion.

As previously mentioned in \secref{sub_sec_host_interfaceing}, once the CAESAR receives the custom ISA, it initially decodes the instructions required for executing a \dnn. The \dnn parameters are established with the activation of the $DNN\_start$ signal. Multiple flag signals, such as those for task allocation and scheduler status indicating idle and busy states, along with an error handling flag, ensure the entire system is informed of any errors at any stage, allowing CAESAR to take appropriate actions. A pipeline coordination flag manages data flow, preventing data stalling while monitoring dynamic parameters. Debugging and optimization monitors utilize these flags to determine fine-tuning requirements or to identify design bottlenecks. Additionally, two primary status flags are included: the Idle FLAG, which indicates whether sub-blocks or the entire SYCore are in an idle state, and the busy status flag, which informs if a sub-block is actively processing data or if data transfer is underway. 

The ISA register is tasked with storing the active instructions. Conversely, the Scheduler block is charged with selecting sub-block addresses and managing the dataflow \fsm, effectively utilizing quantization, pruning, and sparsity, among other methods. The design's size is chosen by the scheduler for parallel execution, and efficient mapping for large workloads poses a significant challenge. The final component of CAESAR is the Data Flow \fsm, responsible for directing data processing, computing each layer, and signalling with $layer\_done$. Once the entire \dnn completes, the $DNN\_Done$ signal is emitted, and all outputs are stored in the L2 shared cache. A data management block oversees all data transfers between the SYCore and the L2 cache, a task included in the Address Mapper, ensuring data localization and storage.

\begin{figure}
    \centering
    \includegraphics[width=\linewidth]{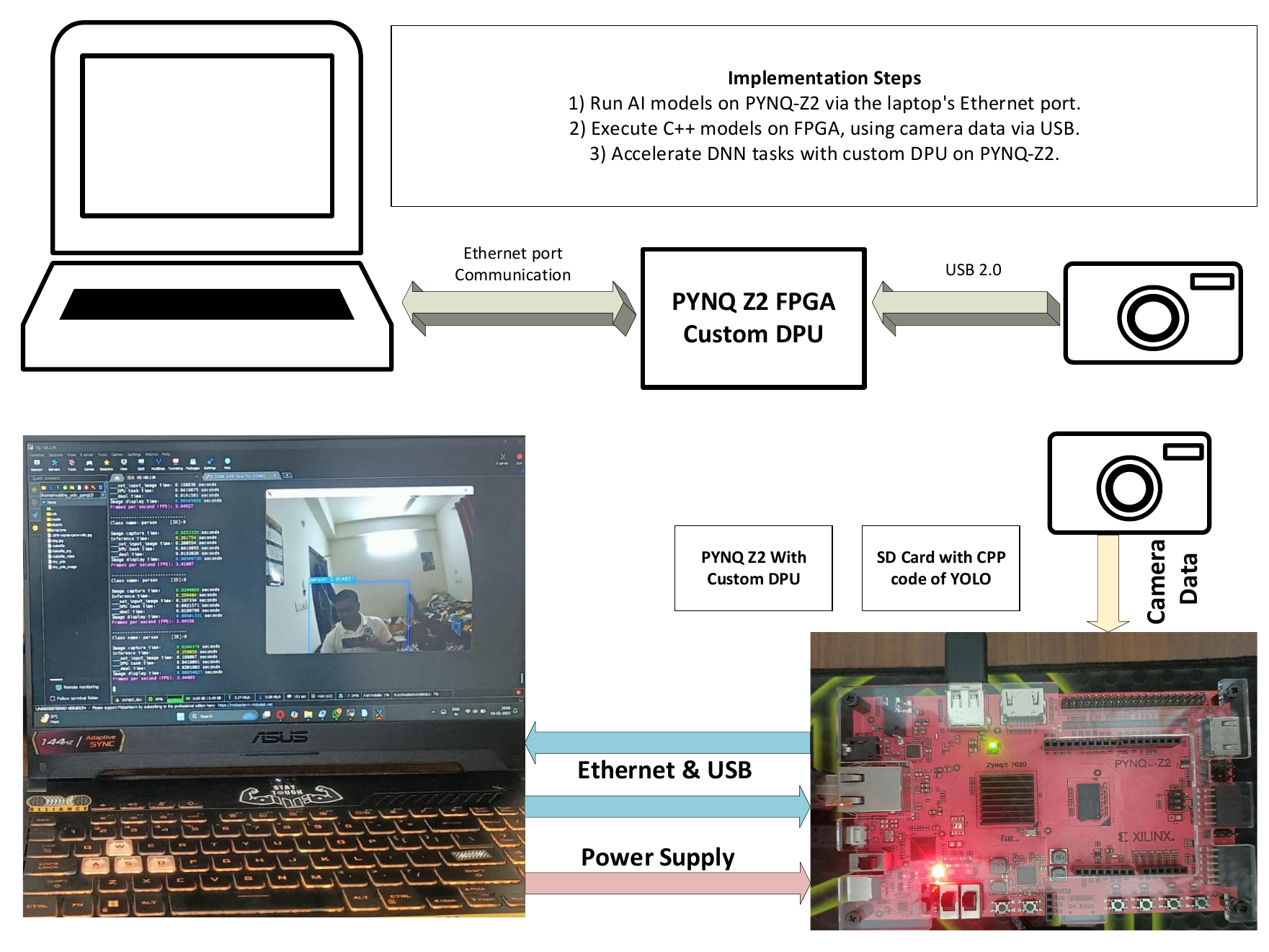}
    \caption{Pynq Z2 Custom DPU Prototype Implementation}
    \label{prototype}
\end{figure}

\subsection{Dataflow and Functioning of the array}
\label{subsec_Dataflow and Functioning of the array}

To comprehend the data and control flow of the CAESAR controller in conjunction with the SYCore, one must gain a clearer grasp of the data flow, particularly as outlined in \secref{VGG16-CIFAR100_SA-mapping}, where the data mapping is described according to computational demands. As previously mentioned, instructions for data parameters initialize the \dnn sizes within the \dnn parameter registers, upon which the scheduler organizes the computations executed by the SYCore. The process begins with mapping the initial Convolutional layer. In column 2, Layer Specifications ($k$\X{$k$}) denote the kernel dimension by $C_\text{in}\times C_\text{out}\times (H\times W)$, correlating to the input image size. In the first convolution, we observe a kernel size of 3\X{3} and an input size of 32\X{32}, with three input channels ($C_\text{in}$) due to RGB, and 64 output channels ($C_\text{out}$). Given the input size of 32\X{32}, these can be efficiently allocated across the SYCore, which is composed of 32\X{32} RPEs, resulting in a total of 1728 MAC operations. The Op Cycles incorporate 1728+4 clock cycles, accounting for each \mac computation, plus additional cycles initially required by the RPE. This design optimizes performance, achieving a complete execution time of 2.65\us, with a power consumption of 0.63\mW. We now proceed to analyze this flow with greater depth.

Following the storage of the \dnn parameter, the scheduler evaluates the initial design to implement it and observes that the input data size completely occupies the Systolic array by utilizing all the Sub-blocks. When the SYCore reaches full capacity, memory access to the L2 cache is facilitated through the AXI interconnect. In this scenario, since no sparsity or pruning is applied that would modify data handling, an address mapper generates a sparse data format by assigning the appropriate addresses to prevent any data issues, which is best understood with an example. Suppose pruning occurs at a 4:9 ratio, specifically weight pruning, meaning some weights are set to zero. This implies that no data is present where the smallest weight values are zeros, removing them from computation to avoid multiplying by zero, known as sparsity. In this design, such a technique is applied at the processor level, altering data patterns which must be recognized by the address mapper. Furthermore, this requires documentation in the \dnn parameters to achieve a more efficient scheduling algorithm and optimize resource usage. By examining various convolution layers where each Sub-Block operates on a 4\X{4} architecture, input mapping at 4\X{4} achieves 100\% utilization; however, efficiency declines as the input decreases in size. The hardware also incorporates tiling to effectively manage the solution, and a complex data reordering flow can control this issue. Additionally, the design relies on the processor to conduct the max-pooling operation. 

Moreover, the integration of the RPE significantly enhances performance. Once the data is distributed into their corresponding sub-blocks, the data Fetcher retrieves all the mapped data from these specified sub-blocks. Subsequently, the \dnn executions commence, with Flag variables monitoring the sub-blocks and activating the DNN\_start and Layer\_start signals to enable them. For each clock cycle, the subsequent inputs and weights are provided to the data, facilitating the execution of all multiplication operations. The output begins arriving after 45 clock cycles, once the initial sub-block RPEs are stored in the SYCore's L1 cache, which is necessary for the next layer's input. Upon complete data reception, the Layer\_done signal is activated, transitioning the RPE state to IDLE. This process repeats until it reaches the initial pooling layer, where output data is forwarded to the L2 cache via the RISC-V processor for pooling operations. All tasks proceed similarly. In convolutional operations, the Data Flow \fsm unit iterates the RPEs across sub-blocks. When the \dnn processes the Flatten layer, data is transformed into a one-dimensional format using SYCore, with adjustments to address and data mapping, after which the FC layers compute the input size multiplied by 1024 SYCore computations, thus again achieving full utilization of SYCore in the final FC as the requirement is 256 for SYCore consumption. A 40\% utilization of CAESAR is observed. Consequently, the total execution time resulted in a delay of 148.6 \ms per execution of an image. Thus, SYCore with a CEASER control engine can achieve up to a maximum of 29.68 inference frames per joule.

\section{Evaluation}
\label{Evaluation Pages}

In the assessment of hardware parameters for optimizing \dnn performance, various design strategies such as fully parallel, sparse, layer-reuse, and our proposed bit-width quantization approach have been applied. The fully parallel \dnn configuration boasts significant throughput; however, it incurs considerable area overhead, thus elevating power consumption. To enhance design optimization, even at the \mac level, we have developed a refined \PE design, comparing \mac results of our proposed design, as documented in \tblref{tab:MAC-ASIC-result}. This table outlines foundational results from different \mac configurations, all computed using CMOS 28nm HPC+ technology and a frequency of 100 MHz. It contrasts metrics like Area, Power, Delay, ADP, Energy, and Priority, focusing on \mac/PE levels, subsequently analyzed for architectural efficiency in terms of compute density (TOPS/mm$^2$) and energy efficiency (TOPS/W), assessed for a 32\X{32} Systolic array. We observe an energy efficiency enhancement of up to 96.7\% at a frequency of 100\MHz. The proposed design's pipelining allows operations at frequencies as high as 3\GHz, demonstrating a 92.15\% improvement in compute density over~\cite{Waris}. Given that \mac hardware is crucial for power savings in AI applications, the proposed design is significant, achieving 31.28 TOPS/W in energy efficiency. At FPGA levels, \mac achieves notable results in terms of 17 LUTs and 26 registers for resource utilization, offering a delay of 1.46 \us at total power consumption of 0.54 \uW. The power-delay product serves as a robust comparative metric due to substantial efficiency improvements introduced by the pipelined \cordic, which enables a more efficient \mac operation at high frequencies.

\begin{table*}[]
    \caption{FPGA \mac comparison}
    \label{tab:MAC-FPGA-Results}
    \resizebox{0.8\textwidth}{!}{%
    \begin{tabular}{ccccccccc}
        \hline
        \multicolumn{1}{c|}{MAC} & \multicolumn{1}{c|}{\textbf{Xilinx IP~\cite{Xilinx-IP}}} & \multicolumn{1}{c|}{\textbf{Vedic \mac~\cite{raut2020cordic}}} & \multicolumn{1}{c|}{\textbf{CORDIC-MAC~\cite{raut2023empirical-ACM}}} & \multicolumn{1}{c|}{\textbf{Wallace \mac~\cite{liu2020systolic}}} & \multicolumn{1}{c|}{\textbf{PipeMAC~\cite{raut2023designingperformacecentricMAC}}} & \multicolumn{1}{c|}{\textbf{PNE \mac~\cite{FP4_8-for-High-Precision-FPGA-CNN}}} & \multicolumn{1}{c|}{\textbf{Booth \mac~\cite{Ratko2}}} & \textbf{Proposed} \\ \hline
        
        \multicolumn{1}{c|}{LUT} & \multicolumn{1}{c|}{53} & \multicolumn{1}{c|}{159} & \multicolumn{1}{c|}{35} & \multicolumn{1}{c|}{105} & \multicolumn{1}{c|}{23} & \multicolumn{1}{c|}{46} & \multicolumn{1}{c|}{83} & 17 \\ \hline
        \multicolumn{1}{c|}{Reg} & \multicolumn{1}{c|}{28} & \multicolumn{1}{c|}{245} & \multicolumn{1}{c|}{58} & \multicolumn{1}{c|}{112} & \multicolumn{1}{c|}{22} & \multicolumn{1}{c|}{20} & \multicolumn{1}{c|}{61} & 26 \\ \hline
        \multicolumn{9}{c}{Delay (us)} \\ \hline
        \multicolumn{1}{c|}{Logic} & \multicolumn{1}{c|}{0.813} & \multicolumn{1}{c|}{\multirow{2}{*}{4.48}} & \multicolumn{1}{c|}{0.713} & \multicolumn{1}{c|}{\multirow{2}{*}{2.59}} & \multicolumn{1}{c|}{\multirow{2}{*}{1.86}} & \multicolumn{1}{c|}{0.38} & \multicolumn{1}{c|}{\multirow{2}{*}{3.08}} & \multirow{2}{*}{1.43} \\ \cline{1-2} \cline{4-4} \cline{7-7}
        \multicolumn{1}{c|}{Signal} & \multicolumn{1}{c|}{2.277} & \multicolumn{1}{c|}{} & \multicolumn{1}{c|}{0.692} & \multicolumn{1}{c|}{} & \multicolumn{1}{c|}{} & \multicolumn{1}{c|}{0.87} & \multicolumn{1}{c|}{} &  \\ \hline
        \multicolumn{9}{c}{Power (uW)} \\ \hline
        \multicolumn{1}{c|}{Logic} & \multicolumn{1}{c|}{0.27} & \multicolumn{1}{c|}{\multirow{3}{*}{1.31}} & \multicolumn{1}{c|}{0.16} & \multicolumn{1}{c|}{\multirow{3}{*}{1.21}} & \multicolumn{1}{c|}{\multirow{3}{*}{0.22}} & \multicolumn{1}{c|}{0.18} & \multicolumn{1}{c|}{\multirow{3}{*}{0.9}} & 0.09 \\ \cline{1-2} \cline{4-4} \cline{7-7} \cline{9-9} 
        \multicolumn{1}{c|}{Signal} & \multicolumn{1}{c|}{0.21} & \multicolumn{1}{c|}{} & \multicolumn{1}{c|}{0.2} & \multicolumn{1}{c|}{} & \multicolumn{1}{c|}{} & \multicolumn{1}{c|}{0.24} & \multicolumn{1}{c|}{} & 0.16 \\ \cline{1-2} \cline{4-4} \cline{7-7} \cline{9-9} 
        \multicolumn{1}{c|}{Dynamic} & \multicolumn{1}{c|}{3} & \multicolumn{1}{c|}{} & \multicolumn{1}{c|}{4} & \multicolumn{1}{c|}{} & \multicolumn{1}{c|}{} & \multicolumn{1}{c|}{2} & \multicolumn{1}{c|}{} & 0.34 \\ \hline
        \multicolumn{1}{c|}{PDP} & \multicolumn{1}{c|}{9.57} & \multicolumn{1}{c|}{5.86} & \multicolumn{1}{c|}{5.62} & \multicolumn{1}{c|}{3.13} & \multicolumn{1}{c|}{0.4092} & \multicolumn{1}{c|}{2.52} & \multicolumn{1}{c|}{2.77} & 0.86 \\ \hline
    \end{tabular}}
\end{table*}

\begin{table*}
    \centering
    \caption{ASIC \mac comparison on 28 nm}
    \label{tab:MAC-ASIC-result}
    \resizebox{0.8\linewidth}{!}{%
    \begin{tabular}{l|ccc|cc|cc|cc} 
        \hline
        \textbf{Parameter} & \textbf{Area (µm²)} & \textbf{Total Power (µW)} & \textbf{Delay (ns)} & \textbf{Energy (pJ)} & \textbf{Area-Delay-Power (ADP)} & \textbf{TOPS/mm²} & \textbf{Relative (\%)} & \textbf{TOPS/W} & \textbf{Relative (\%)} \\ 
        \hline
        \textbf{Liu et al.~\cite{LiuTCASI}} & 838 & 502 & 6.44 & 3.23 & 2.71E+06 & 1.32 & 8.32 & 2.38 & 7.61 \\ 
        \hline
        \textbf{Ashar et al.~\cite{ashar2024quantmac}} & 501 & 122 & 4.27 & 0.52 & 2.61E+05 & 3.34 & 20.99 & 14.77 & 47.21 \\ 
        \hline
        \textbf{Raut et al.~\cite{raut2023empirical-ACM}} & 307 & 144 & 3.62 & 0.52 & 1.60E+05 & 6.43 & 40.41 & 6.92 & 22.13 \\ 
        \hline
        \textbf{Ratko et al.~\cite{Ratko}} & 272 & 195.24 & 2.52 & 0.49 & 1.34E+05 & 10.42 & 65.52 & 11.22 & 35.88 \\ 
        \hline
        \textbf{Ansari et al.~\cite{Ansari}} & 368 & 260.38 & 2.59 & 0.67 & 2.48E+05 & 7.49 & 47.12 & 8.07 & 25.80 \\ 
        \hline
        \textbf{Yin et al.~\cite{yin2023blocksparse}} & 244 & 227.4 & 2.46 & 0.56 & 1.36E+05 & 11.90 & 74.82 & 12.82 & 40.98 \\ 
        \hline
        \textbf{Waris et al.~\cite{Waris}} & 771 & 153 & 7.42 & 1.14 & 8.75E+05 & 1.25 & 7.85 & 1.34 & 4.30 \\ 
        \hline
        \textbf{Warris et al.~\cite{Warris}} & 606 & 552.06 & 2.47 & 1.36 & 8.26E+05 & 4.77 & 30.00 & 5.14 & 16.43 \\ 
        \hline
        \textbf{Ratko et al.~\cite{Ratko}} & 440 & 200.34 & 2.88 & 0.58 & 2.54E+05 & 5.64 & 35.44 & 6.07 & 19.41 \\ 
        \hline
        \textbf{Kim et al.~\cite{Kim}} & 297 & 211.64 & 2.71 & 0.57 & 1.70E+05 & 8.87 & 55.80 & 9.56 & 30.56 \\ 
        \hline
        \textbf{Leone et al.~\cite{Leone}} & 373 & 283.44 & 2.93 & 0.83 & 3.10E+05 & 6.54 & 41.09 & 7.04 & 22.50 \\ 
        \hline
        \textbf{Liu et al.~\cite{Liuy2}} & 294 & 224.45 & 2.36 & 0.53 & 1.56E+05 & 10.29 & 64.73 & 11.09 & 35.45 \\ 
        \hline
        \textbf{Reza et al.~\cite{Reza}} & 447 & 199.5 & 2.54 & 0.51 & 2.27E+05 & 6.29 & 39.56 & 6.78 & 21.66 \\ 
        \hline
        \textbf{Liu et al.~\cite{liu2020systolic}} & 815 & 291.66 & 2.93 & 0.85 & 6.96E+05 & 2.99 & 18.81 & 9.00 & 28.78 \\ 
        \hline
        \textbf{Proposed} & 200.5 & 109.8 & 2.24 & 0.25 & 4.93E+04 & 15.90 & 100.00 & 31.28 & 100.00 \\
        \hline
    \end{tabular}
    }
\end{table*}

\begin{table*}[!t]
    \caption{ASIC \mac comparison on 7 nm}
    \label{asic-7mac}
    \resizebox{0.8\linewidth}{!}{%
    \begin{tabular}{l|ccc|cc|cc|cc}
        \hline
        \textbf{Parameter} & \textbf{Area (µm²)} & \textbf{Total Power (µW)} & \textbf{Delay (ns)} & \textbf{Energy (pJ)} & \textbf{Area-Delay-Power (ADP)} & \textbf{TOPS/mm²} & \textbf{Relative (\%)} & \textbf{TOPS/W} & \textbf{Relative (\%)} \\ \hline
        \textbf{Liu et al.~\cite{LiuTCASI}} & 273 & 188 & 5.14 & 0.97 & 2.64E+05 & 5.09 & 106.2 & 7.96 & 150.6 \\ \hline
        \textbf{Ashar et al.~\cite{ashar2024quantmac}} & 161 & 45.8 & 3.38 & 0.15 & 2.49E+04 & 13.13 & 273.8 & 49.69 & 940.4 \\ \hline
        \textbf{Raut et al.~\cite{raut2023empirical-ACM}} & 99 & 54 & 2.86 & 0.15 & 1.53E+04 & 25.23 & 526.2 & 27.17 & 514.1 \\ \hline
        \textbf{Ratko et al.~\cite{Ratko}} & 255 & 65 & 2.78 & 0.18 & 4.61E+04 & 10.08 & 210.2 & 10.85 & 205.4 \\ \hline
        \textbf{Ansari et al.~\cite{Ansari}} & 354 & 81 & 2.04 & 0.17 & 5.85E+04 & 9.89 & 206.3 & 10.65 & 201.6 \\ \hline
        \textbf{Yin et al.~\cite{yin2023blocksparse}} & 234 & 70.4 & 1.94 & 0.14 & 3.20E+04 & 15.73 & 328.2 & 16.94 & 320.7 \\ \hline
        \textbf{Waris et al.~\cite{Waris}} & 248 & 57.3 & 5.87 & 0.34 & 8.34E+04 & 4.91 & 102.3 & 5.28 & 100.0 \\ \hline
        \textbf{Warris et al.~\cite{Warris}} & 472 & 96.4 & 1.96 & 0.19 & 8.92E+04 & 7.72 & 161.1 & 8.31 & 157.4 \\ \hline
        \textbf{Ratko et al.~\cite{Ratko}} & 324 & 68 & 2.28 & 0.16 & 5.02E+04 & 9.67 & 201.7 & 10.41 & 197.1 \\ \hline
        \textbf{Kim et al.~\cite{Kim}} & 299 & 87 & 2.13 & 0.19 & 5.54E+04 & 11.22 & 234.0 & 12.08 & 228.6 \\ \hline
        \textbf{Leone et al.~\cite{Leone}} & 267 & 76 & 3.19 & 0.24 & 6.47E+04 & 8.39 & 174.9 & 9.03 & 170.9 \\ \hline
        \textbf{Liu et al.~\cite{Liuy2}} & 246 & 95 & 1.86 & 0.18 & 4.35E+04 & 15.61 & 325.6 & 16.81 & 318.2 \\ \hline
        \textbf{Reza et al.~\cite{Reza}} & 437 & 113 & 2.09 & 0.24 & 1.03E+05 & 7.82 & 163.1 & 8.42 & 159.4 \\ \hline
        \textbf{Liu et al.~\cite{liu2020systolic}} & 386 & 102 & 3.86 & 0.39 & 1.52E+05 & 4.79 & 100.0 & 19.54 & 369.7 \\ \hline
        \textbf{Proposed} & 64.6 & 41 & 1.77 & 0.07 & 4.69E+03 & 62.47 & 1303.1 & 106.00 & 2006.0 \\ \hline
    \end{tabular}}
\end{table*}

\subsection{Experimental Setup and Validation}
\label{subsec_Experimental Setup and Validation}
For Validation and Setup, we executed tests on the design. Initially, we focused on testing the \cordic by employing Python on Google COLAB, which allowed us to determine the \cordic requirements for implementing the necessary number of stages for operations like \mac, $\tanh$, Sigmoid, and \softmax, as illustrated in \figref{fig:sigmoid_error_plot1}, \figref{fig:tanh_error_plot}, and \figref{fig:SoftMax_error_plot}. The error analysis is detailed in \secref{subsubsec_Pareto Analysis}, based on various error types discussed. It was found that conducting five iterations is a reasonable approach for \cordic computations. We then proceeded to the design phase of the RPE. We created RTL code for both single and multi-staged reconfigurable processing engines, primarily utilizing \cordic. Using this approach, we gathered ASIC results for the design using Cadence Genus in a TSMC 28\nm technology node at a standard process corner. As shown in the table, we scripted the Tcl for the design with a clock period of 10\ns. Further scaling of the technology node can yield improved results. Prior to this, we also conducted RPE Emulation using Vivado 2024.1 software. Correct output values were achieved after adjusting select signals and proper clock mapping; following optimized synthesis results, we proceeded to the implementation and integration of the systolic array, where the design of the control engine and computational metrics such as Excel TOPS/mm$^2$ and TOPS/W were calculated.

\subsection{Software-Based Evaluation and Validation of Inference Accuracy}
\label{subsec_Software-Based Evaluation and Validation of Inference Accuracy}
\begin{figure}
    \centering
    \includegraphics[width=0.8\linewidth]{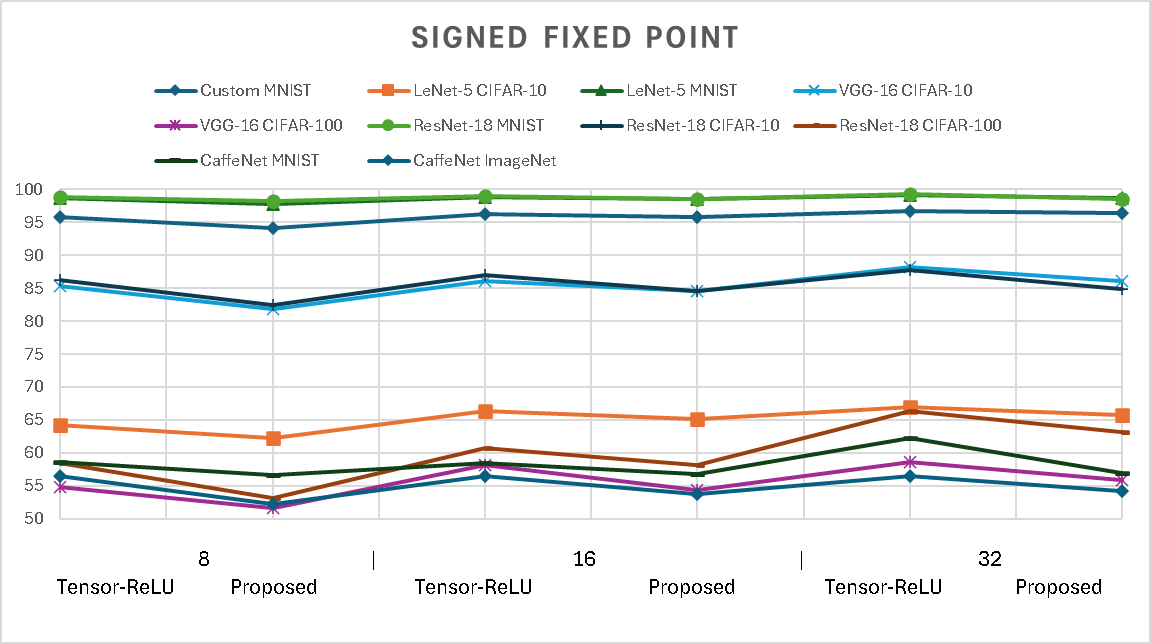}
    \caption{Accuracy at Several Deep Neural Nets}
    \label{fig:accuracy-plot-result}
\end{figure}

The models' accuracy is depicted in \figref{fig:accuracy-plot-result}, indicating minimal variation across the different models. Notably, there is a minor decrease in accuracy attributable to the \cordic approximation errors. Our analysis involved AI models, including custom models, VGG-16 on CIFAR-100, CaffeNet, and ResNet, with LeNet-5 evaluated on MNIST and CIFAR-10. The MNIST dataset is relatively small and consists of handwritten digits, while CIFAR-10 is a larger dataset used for image classification and is included in certain complex datasets. ImageNet, meanwhile, is an extensive dataset used for object classification and prediction in video feeds. We considered a diverse range of models, with LeNet-5 exemplifying smaller architectures, while larger models tend to attain slightly reduced accuracy that remains comparable to \sota achievements. This evaluation also encompassed Transformers at different bit precisions. When juxtaposed with other tensor-based models \cite{Tensor-Std}, the proposed solution shows less than a 2\% decrease in accuracy. Accuracy results were obtained through TensorFlow using the Qkeras library to adjust precision, as illustrated in \figref{fig:accuracy-plot-result}, which assessed data widths of 8, 16, and 32 bits. This underscores accuracy improvements achievable through various bit precisions. Employing integer/fixed-point data formats suggests favouring lower bit widths due to the negligible accuracy gains relative to their cost. Consequently, training loops were executed for 50-80 epochs to attain optimal accuracy. Moreover, it was found that pruning up to 40\% did not incur any per-layer loss, though further pruning is feasible for certain layers. Our architecture supports hardware execution using CAESER.

\subsection{Pareto Analysis of RPE Design Parameters: Pipeline Stages, Bit-Precision, and Integer Bits, etc. other design parameters (MAC level comparison and Pareto)}
\label{sub_sec_Pareto Analysis of RPE Design Parameters: Pipeline Stages, Bit-Precision, and Integer Bits, etc. other design parameters (MAC level comparison and Pareto)}
The development and validation of a \cordic \PE with adjustable \cordic \mac and \AF units is conducted using a Python-based emulator to evaluate both accuracy and hardware performance. Pareto analysis is utilized to identify the ideal number of \cordic stages that optimally balance accuracy, area, power consumption, and delay. In the context of \mac design, this achieves enhanced performance throughput per watt, albeit with a compromise in accuracy, as evidenced by results on the MNIST dataset. Specifically, for an optimized iteration count, accuracy loss is noted in the \mac with five stages, with a normalized mean error of $6.31 \times 10^{-5}$. This corresponds to a normalized mean error distance of $0.00783$, a mean relative error distance of $0.02675$, and a normalized maximum error distance of $0.03131$, indicating minimal error in \mac operations. Based on these metrics, it is concluded that the proposed \mac design offers approximately 23.34\X improvement in throughput per watt and up to 12.72\X improvement in computational density compared to the \sota benchmarks, calculated by dividing throughput by total power and area, respectively. This performance advantage is evident despite processing 8-bit data in FxP format, as the pipelined \mac operation produces one multiplication per clock cycle, allowing the design to reach a frequency up to 3\GHz.

The findings summarized in \tblref{tab:HW-arch_result} concern a system functioning at 100\MHz, utilizing a Systolic Array made up of 32\X{32} RPEs, thus comprising up to 1024 RPEs in total. The RPE flexibility enables various AI operations on the Accelerator, including \RNNs, which handle sequential input processing within this framework. When comparing our \mac architecture with existing models and assuming our system realizes 100\% efficiency, it becomes apparent that other accelerators, such as those by Yin et al.~\cite{yin2023blocksparse}, achieve up to 74.82\% efficiency, which remains approximately 25\% less efficient than our design. Furthermore, Liu et al.~\cite{LiuTCASI} achieved 64.73\% and Ratko et al.~\cite{Ratko} reported 65.52\%, making them the nearest competitors in efficiency. The \mac not only refines computational efficiency but also boosts \AF performance, as displayed in the Pareto analysis for conventional \dnn \AFs like sigmoid, $\tanh$, and \relu in \dnns and \RNNs, seen in Fig. \ref{fig:tanh_error_plot}, and \ref{fig:SoftMax_error_plot}. Additionally, the design accommodates Transformers by supporting \softmax, albeit requiring added clock cycles due to expansive probability calculations. 

Beyond classification, the hardware configuration, optimized for five iterations, minimizes error with a precision reduction to 8-bit. Increased precision leads to heightened error, while \relu remains error-free irrespective of iteration count, executing within a single clock cycle, similar to other \AFs. This architecture supports operation at a frequency of 3\GHz, needing nine clock cycles—five for hyperbolic functions and four for division calculations. Enhanced sparsity and pruning further improve throughput, reducing latency by 1.7\X, \mac utilization by 2\X, and computational parameters by 1.8\X through commercial 4:9 pruning methods. The ASIC evaluations also cover aspect, power, and delay, with concurrent FPGA analysis in \tblref{tab:HW-arch_result}, respectively. The process incurs a 148\ms delay and a total power of 1.524\W at a 100\MHz operation frequency.

\subsection{Hardware Implementation and Comparative Performance Analysis (Architectural Level of analysis)}

\begin{table*}[]
    \caption{Hardware Implementation Report with Proposed Systolic Array Architecture and State-of-the-Arts \dnn Designs}
    \label{tab:HW-arch_result}
    \resizebox{0.8\textwidth}{!}{%
    \begin{tabular}{l|ccc|cccc|cc}
        \hline
        \textbf{} & \textbf{Platform} & \textbf{Model} & \textbf{Precision} & \textbf{\begin{tabular}[c]{@{}c@{}}LUTs \\ (Thousands)\end{tabular}} & \textbf{\begin{tabular}[c]{@{}c@{}}Registers\\  (Thousands)\end{tabular}} & \textbf{DSPs} & \textbf{\begin{tabular}[c]{@{}c@{}}Op. Freq \\ (MHz)\end{tabular}} & \textbf{\begin{tabular}[c]{@{}c@{}}Energy efficiency \\ (GOPS/W)\end{tabular}} & \textbf{\begin{tabular}[c]{@{}c@{}}Power\\  (Watts)\end{tabular}} \\ \hline
        \textbf{Cong et. al~\cite{Cong}} & Pynq-Z1 & Custom & N/A & 44 & 40 & 200 & 100 & N/A & 2.3 \\
        \textbf{Yifan et. al~\cite{Yifan}} & ZU3EG & DiracDeltaNet & 1A4W & 24 & 30 & 37 & 250 & 8.5 & 5.5 \\
        \textbf{Bi et. al~\cite{Bi}} & ZU3EG & ResNet-50 & 8 & 41 & 45 & 250 & 150 & 45 & 1.5 \\
        \textbf{Xiaodi et. al~\cite{yin2023blocksparse}} & ZCU102 & MobileNetV2 & 16 & 195 & 94 & 880 & 190 & N/A & 13.4 \\
        \textbf{Liqiang et. al~\cite{Liqiang}} & ZCU102 & VGG16 & 16 & 130 & 69 & 365 & 200 & 12 & 23.5 \\
        \textbf{Alessandro et. al~\cite{Alessandro}} & Zynq7 & VGG16 & 16 & 230 & 110 & 130 & 60 & 27 & 1.1 \\
        \textbf{Adithya et. al~\cite{RAMAN-IoTJ24}} & Ti60 & MobileNetV1 & 8 & 38 & 8.75 & 60 & 75 & 95 & 0.24 \\
        \textbf{Xiaoru et al.~\cite{Xiaoru}} & Arria10 & MobileNetV2 & 8 & 100 & N/A & 510 & 175 & 19 & 4.6 \\
        \textbf{Gopal et. al~\cite{raut2023empirical-ACM}} & ZC706 & Custom & 8 & 115 & 115 & 32 & 100 & 4.5 & 2 \\
        \textbf{Bradley et.al~\cite{Bradley}} & VC707 & Custom & N/A & 240 & 200 & 110 & 170 & N/A & 2.2 \\
        \textbf{Proposed} & VC707 & VGG16 & 8 & 91 & 126 & 1350 & 466 & 298 & 1.53 \\ \hline
    \end{tabular}%
    }
\end{table*}

Building on the proposition of the SYCore architecture, we proceeded to conduct a detailed analysis of its implementation on FPGA. Specifically, we assessed the VC707 Virtex-7 board, which operates at 8-bit precision and a built-in clock frequency of 500 MHz. The architecture demonstrated an energy efficiency of 298 GOPS/W. Regarding resource utilization, the architecture exhibited demands of 91k LUT, 126k registers, and 1350 DSPs. This marks as improved of approximately 1.5 $\times$, 2 $\times$ and 2.5 2 $\times$ over ]\cite{Liqiang}, \cite{Xiaoru}, and \cite{Alessandro, Bradley} respectively. Furthermore, utilizing the Systolic array in conjunction with the RPE design yielded a throughput of up to 455 GOPS at 853 MHz frequency within FPGAs. Comparatively, Bradley et al.~\cite{Bradley} achieved a resource allocation of 240k LUTs, 200k registers, and 110 DSPs on similar hardware, which is substantially resource-intensive for any FPGA. Our design achieves the highest energy efficiency of 298 GOPS/W over SoTA works. The proposed architecture enables operation at higher frequencies for enhanced throughput. Additionally, power efficiency is critical, particularly for edge devices with limited power availability. In this context, the proposed architecture supports the implementation of complex classification models like VGG-16 on the hardware.

\subsection{Evaluating Hardware Parameters for Optimal \dnn Performance}

\begin{figure}[h]
    \centering
    \begin{minipage}{0.49\textwidth}
        \centering
        \includegraphics[width=\linewidth, height = 40mm]{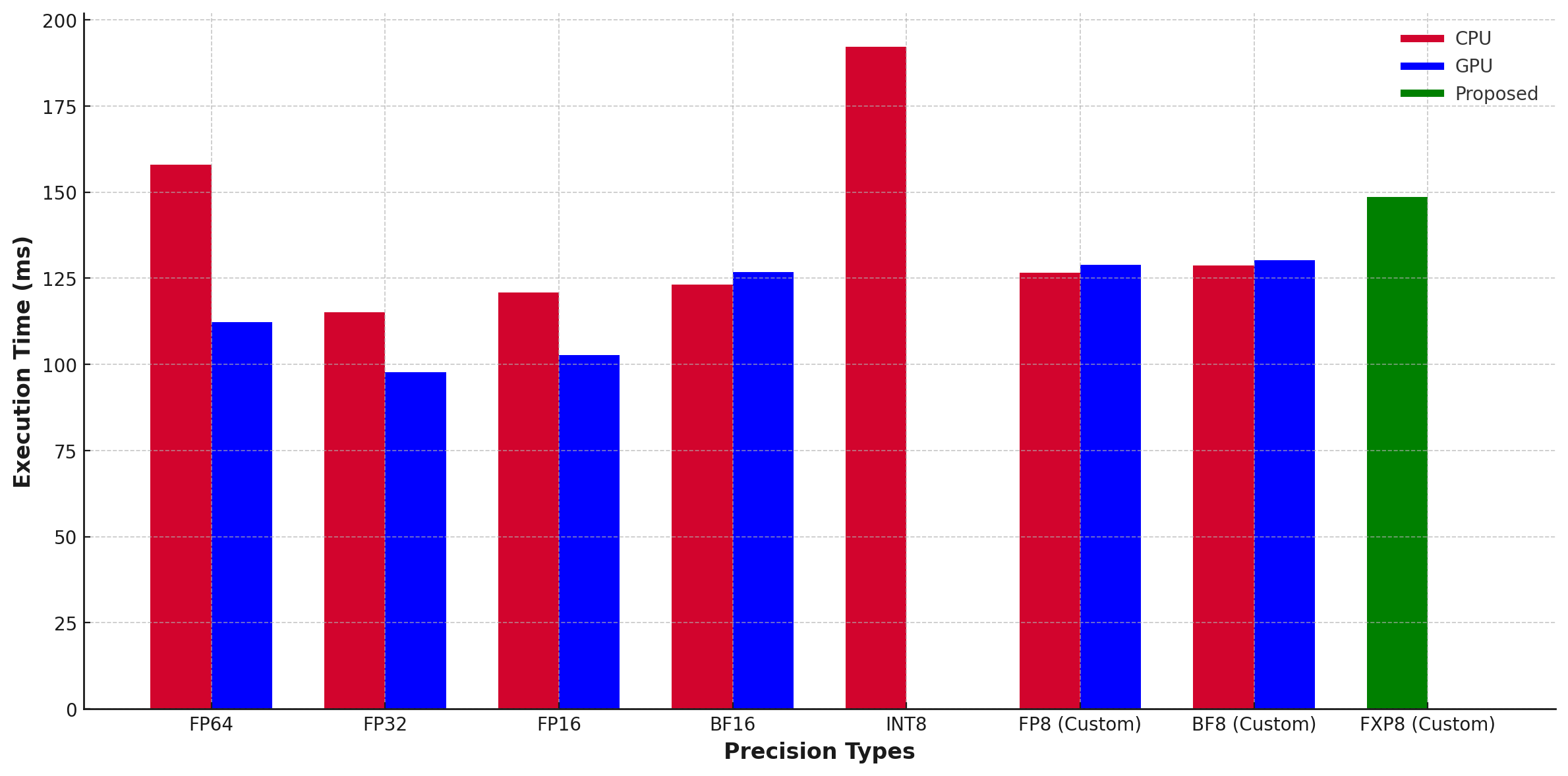}
        \caption{Execution time for different precision at CPU and GPU platforms compared to proposed model}
        \label{fig:runtime-result}
    \end{minipage}
    \hfill
    \begin{minipage}{0.49\textwidth}
        \centering
        \includegraphics[width=\linewidth, height = 40mm]{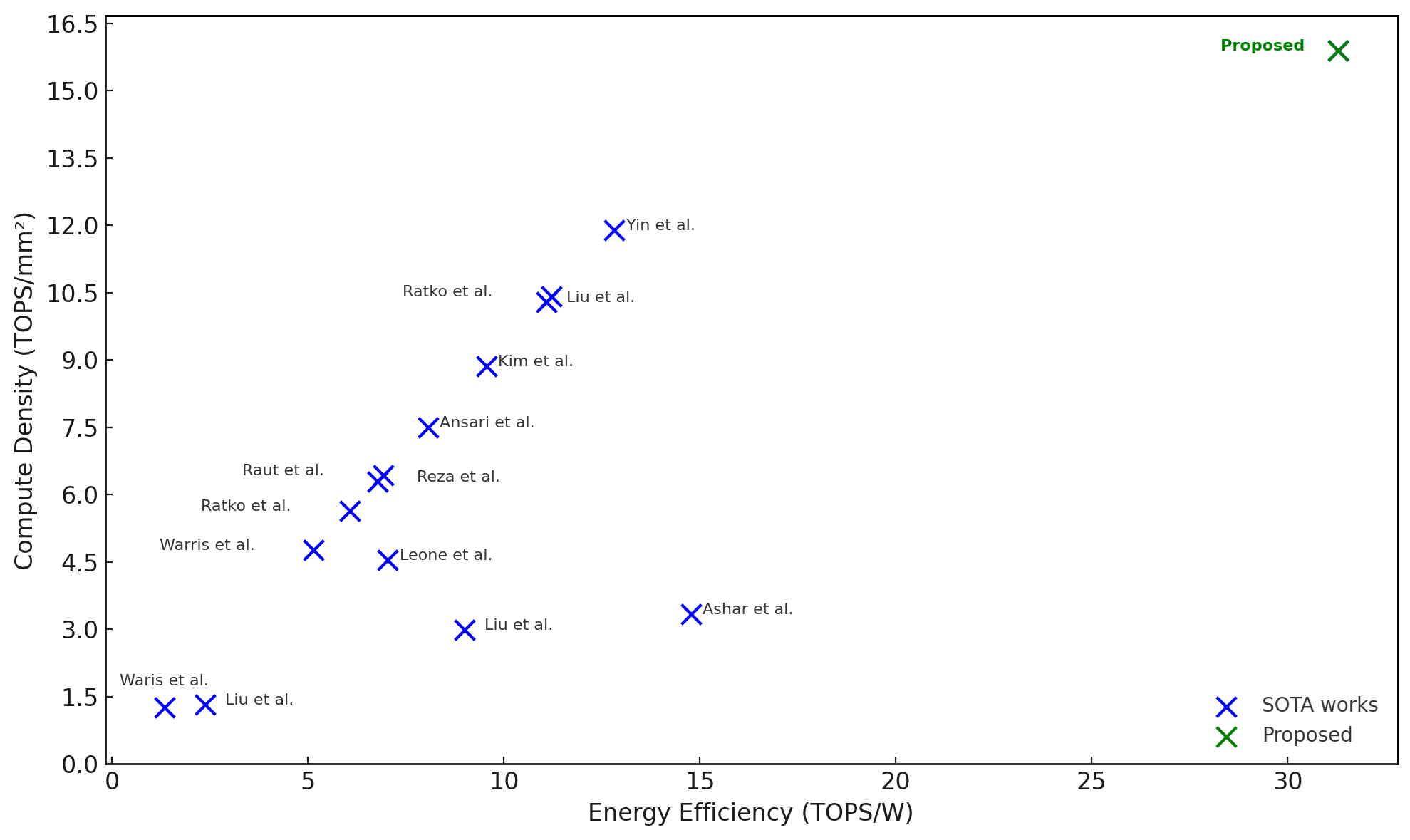}
        \caption{Performance analysis at CMOS 28 nm and comparison with \sota works}
        \label{fig:performance_comparision_efficiency}
    \end{minipage}
    
\end{figure}

Post-APR CMOS 28\,nm HPC+ and 45\,nm with Synopsys Design Compiler results for DA-VINCI design with the SOTA comparison is reported in Table \ref{tab:ASIC-comp}, marking reduced area up to 16.2$\times$ and 4$\times$ while optimizing the critical delay to 7.8$\times$ and 1.3$\times$ compared to the SoTA works~\cite{Softmax-DNNTraining} and~\cite{RECON-CORDICNeuron} respectively, while 2.1$\times$ and 14.3$\times$ power reduction than ~\cite{CORDICAF-LSTM} and~\cite{RECON-CORDICNeuron}. It also shows improvement up to 11.5$\times$ area, 1.75$\times$ critical delay, and 17.3$\times$ power consumption compared to ReCON~\cite{RECON-CORDICNeuron}, while reducing 2.5$\times$ power and 1.8$\times$ area compared to ~\cite{CORDICAF-LSTM} and~\cite{Softmax-taylor-DNN} with CMOS 28\,nm. 

\begin{table}
    \caption{FPGA Resources Utilization, compared with prior works}
    \label{tab:FPGA-comp}
    \resizebox{0.75\columnwidth}{!}{
    \begin{tabular}{c|c|c|c|c|c}
    \hline
    \textbf{Design} & \textbf{ReAFM~\cite{ReAFM-NN}} & \textbf{Shen et al.~\cite{CORDICAF-RNN}} & \textbf{AFB~\cite{CORDICAF-LSTM}} & \textbf{AxSF~\cite{Softmax-taylor-DNN}} & \textbf{DA-VINCI} \\ \hline
    \textbf{AF} & sigmoid-tanh-swish & tanh-sigmoid & tanh-sigmoid & softmax & configurable-AF \\ \hline
    \textbf{Precision} & 12-bit & - & 16-bit & 16-bit & 8/16-bit \\ \hline
    \textbf{Mean error (\%)} & 2.2 & 3.4 & 2.8 & - & 2.78 \\ \hline
    \textbf{FPGA} & Virtex-7 & Zynq-7 & PYNQ-Z1 & Zynq-7 & Zybo \\ \hline
    \textbf{LUTs} & 367 & 2395 & 36286 & 1215 & 137 \\ \hline
    \textbf{FFs} & 298 & 1503 & 24042 & 1012 & 68 \\ \hline
    \textbf{Delay ($\mu$s)} & 350 & 180 & 210 & 332 & 256 \\ \hline
    \textbf{Power (mW)} & - & 0.681 & 125 & 165 & 27.53 \\ \hline
    \end{tabular}
    }
    \vspace{-3mm}
\end{table}

\begin{table}
    \caption{ASIC Resources Utilization, compared with prior works}
    \label{tab:ASIC-comp}
    \resizebox{0.75\columnwidth}{!}{%
    \begin{tabular}{c|c|c|c|c|cc}
    \hline
    \textbf{Design} & \textbf{Zhang et al.~\cite{Softmax-DNNTraining}} & \textbf{AxSF~\cite{Softmax-taylor-DNN}} & \textbf{AFB~\cite{CORDICAF-LSTM}}  & \multicolumn{1}{c|}{\textbf{RECON~\cite{RECON-CORDICNeuron}}} & \multicolumn{2}{c}{\textbf{DA-VINCI}} \\ \hline
    \textbf{AF Support} & softmax & softmax & tanh-sigmoid & \multicolumn{1}{c|}{tanh-sigmoid}  & \multicolumn{2}{c}{configurable-AF} \\ \hline
    \textbf{Precision (bit)} & 32 & 16  & 16 & 16 & \multicolumn{2}{c}{8/16} \\ \hline
    \textbf{Tech. Node (nm)} & 28 & 28 & 45 & \multicolumn{1}{c|}{45} & \multicolumn{1}{c|}{45} & 28 \\ \hline
    \textbf{Area (1000 * $\mu$m\textsuperscript{2})} & 98.78 & 3.82 & 870.5 & 24.61 & \multicolumn{1}{c|}{1.24}& 0.78\\ \hline
    \textbf{Delay (ns)} & 26 & 1.6 & - & 4.76 & \multicolumn{1}{c|}{6.7} & 3.8\\ \hline
    \textbf{Power (mW)} & 24.72 & 1.58 & 151 & 1033 & \multicolumn{1}{c|}{22.3} & 17.5\\ \hline
\end{tabular}}
\vspace{-3mm}
\end{table}

When assessing the complete \dnn for application-level purposes, throughput and latency are critical metrics. In image detection tasks, detection latency directly influences the achievable frames per second for real-time implementation. To illustrate this, we have measured the latency of VGG-16 across different platforms: a CPU, a GPU, and our Proposed design. The CPU used is an AMD Ryzen 7 6800H with Radeon Graphics, operating at a base frequency of 3201 MHz with 16 logical processors manufactured in 7\nm technology. For the GPU comparison, we tested an NVIDIA GeForce RTX 3050 Ti with 4.0\GB memory, consuming 60\W and also fabricated in 7\nm, featuring 80 AI/Tensor cores. Our proposed design, using a single-core configuration at a frequency of 150\MHz, achieves similar performance despite being produced at a 28\nm node. When the SYCore operates multiple instances in parallel with a lower technology node at even higher frequencies, it surpasses conventional CPU and GPU hardware, as demonstrated in \figref{fig:runtime-result}. Our design is evaluated using fixed-point 8-bit (FxP 8-bit) computing, whereas the CPU and GPU are evaluated using FP64, FP32, FP16, FP8, BF16, BF8, and Int8 with the VGG-16 network, with average execution time shown on the y-axis. High latencies, in some cases, stem from a required quantization circuit in the workflow, which converts data to 8-bit integers, introducing delays. Despite this, the proposed solution provides latency comparable to other hardware, with additional flexibility offered by the \AF for overall computation. This enhances our hardware design relative to industry standards.

In conclusion, upon evaluating the throughput efficiency as depicted in \figref{fig:performance_comparision_efficiency}, the newly proposed design exhibits significant advancements in both energy efficiency and Compute Density relative to existing architectures. It achieves an Energy Efficiency exceeding 30 TOPS/W and a Compute Density surpassing 15 TOPS/mm$^2$, whereas comparable architectures offer a compute density not exceeding 13.5 TOPS/mm$^2$ and an Energy Efficiency below 15 TOPS/W.  The detailed implementation approach is demonstrated in Fig. \ref{prototype}. It was noted during utilization analysis that task execution required approximately 150 ms. \tblref{comp-cpa} highlights that the proposed system, when deployed on PYNQ-z2, remains a cost-effective alternative, delivering a comparable latency performance using Memory - 630 KB, and Storage - 16 MB, as previously mentioned. The SYCore, equipped with a CAESAR control engine, is capable of achieving a maximum of 29.68 inference frames per joule.

\begin{table*}[!t]
    \caption{Comparative cost performance analysis with \sota solution}
    \label{comp-cpa}
    \resizebox{0.8\textwidth}{!}{%
    \begin{tabular}{c|c|c|c|c|c}
        \hline
        \textbf{Parameter} & \textbf{Bit-Precision} & \textbf{FPGA \& Cost (\$)} & \textbf{Power (W)} & \textbf{\begin{tabular}[c]{@{}l@{}}MAC Units \\ Utilisation (\%)\end{tabular}} & \textbf{Freq (MHz)} \\ \hline
        \textbf{Proposed} & FXP8 &  PYNQ-Z2,\ 140 &  \textbf{1.524} & {\color[HTML]{38761D} \textbf{83.75+}} & 50 \\ \hline
        \textbf{Kim et al.~\cite{Kim-yolov3}} & 8-bit & A7-100T, 265 & 2.2 & 82.63 & 100 \\ \hline
        \textbf{Jiang et al.~\cite{Jiang}} & 8-bit & ZCU102, 3234 & - & 82.53 & 333 \\ \hline
        \textbf{Trio et al.~\cite{Trio}} & INT8 & ZCU104, 1678 & 1.89 & 71.85 & 187 \\ \hline
        \textbf{Thanh et al.~\cite{DThanh}} & 8-bit & VC707, 5244 & - & 72.70 & 200 \\ \hline
        \textbf{Raut et al.~\cite{raut2023empirical-ACM}} & FXP8 & VC707, 5244 & 1.12 & 54 & 466 \\ \hline
        \textbf{Lee et al.~\cite{WLEE}} & Variable & ZCU102, 3234 & 0.82 & 40.43 & 300 \\ \hline
        \textbf{Subin et al.~\cite{Subin}} & FP8 & XCVU9, 4830 & 5.52 & 34.8 & 150 \\ \hline
    \end{tabular}}
\end{table*}

\section{Conclusion and Future works}

This study presents a reconfigurable processing element (RPE) built on a \cordic platform. This element incorporates five linear \cordic stages tailored for \mac functionalities and additional hyperbolic and division stages to iteratively compute \AFs. The research highlights an optimized architecture for neural network computing that minimizes resource consumption while sustaining high accuracy, as validated through Pareto analysis. The design offers runtime reconfigurability, enabling adaptation of \AFs like sigmoid, $\tanh$, \relu, \softmax based on workload demands. The \cordic hardware has been fine-tuned to curtail area and power expenditures compared to standard designs, making it particularly suitable for resource-constrained environments such as edge computing. A notable aspect of this architecture is its flexibility, accommodating diverse neural network tasks, including \dnns, \RNNs, and Transformers, all within adaptable precision configurations. This yields a scalable design that enhances performance. The findings suggest considerable enhancements in hardware efficiency, with reduced latency and improved throughput in AI processing tasks by a factor of up to 2.5 $\times$. The paper affirms that \cordic-based designs are effective for most AI workloads, especially on edge devices like smartphones, IoT gadgets, robotics, autonomous vehicles, etc. Integrating in-memory design is a promising strategy. We have both validated and prototyped this design on FPGA platforms, specifically the Virtex VC-707 and conducted CMOS 28\nm level evaluations at the \mac level, offering benchmarks against real-world performance outcomes. Additionally, a multi-SYCore architecture supports a multi-tenant workload deployment.

Ultimately, The paper presents a novel, highly reconfigurable architecture based on a \cordic processing engine, capable of functioning as a \mac block or supporting various \AFs such as hyperbolic functions, division, and AI-specific functions like \relu, $\tanh$, sigmoid, and \softmax. The proposed design allows for hardware reconfiguration, making it suitable for various AI models, from feed-forward networks and \dnns to transformers. Additionally, the architecture efficiently exploits sparsity and pruning techniques to improve throughput to 4.57 TOPS while reducing power consumption by 5\X. The system is capable of operating at high frequencies of 3\GHz at the TSMC 28nm technology node. Further scaling will enhance the design, contributing to significant performance gains in AI applications. The provided architecture makes an Edge-compatible design for running various Edge workloads such as LLM, Object detection and classification with minimal delay.




\section*{Acknowledgements}
The authors extend their heartfelt gratitude to the Special Manpower Development Program Chip to Startup, Ministry of Electronics and Information Technology (MeitY), Government of India, for providing the essential research infrastructure that made this work possible. 
They would also like to sincerely thank their colleagues for helpful discussions, including but not limited to Mr Advay Kunte, Ms Akanksha Jain, Mr Akash Sankhe, Mr Astik Sharma, Mr Ayush Awasthi, Mr Dinesh, Mr Giridhar, Ms Komal Gupta, Ms Neha Ashar, Dr Narendra Dhakad, Mr Prabhat Sati, Mr Praneeth, Mr Radheshyam Sharma, Dr Ratko Pilipovic for their invaluable support with technical discussions for this work.
The authors are deeply grateful to the reviewers for their insightful feedback and thorough evaluation, which significantly improved the quality of this manuscript. Any errors that remain are solely the responsibility of the authors.


\ifacmjournal
    \bibliographystyle{ACM-Reference-Format}
\else \ifmicro
    \bibliographystyle{IEEEtranS}
    \fi
\else
    \bibliographystyle{IEEEtran}
\fi

\bibliography{bibliography/final_biblography,bibliography/this_bibliography}




\end{document}